%% file: paper.tex
\documentclass{vldb}
\usepackage{graphicx}
\usepackage{balance}  
\usepackage{booktabs} 
\pdfoutput=1
\usepackage[table]{xcolor}
\usepackage{threeparttable}
\usepackage{footnote}
\usepackage{multirow}
\usepackage{hhline}
\usepackage{diagbox}
\usepackage[ruled,vlined,linesnumbered]{algorithm2e}
\usepackage{graphicx}
\usepackage{tikz}
\usepackage[labelfont=bf]{caption,subfig}
\usepackage{makecell}
\usepackage{hyperref}

\usepackage{amsthm}
\usepackage{mathptmx}
\usepackage{enumitem}
\usepackage[scientific-notation=true]{siunitx}
\setitemize{noitemsep, topsep=0pt, parsep=0pt, partopsep=0pt}
\setlist[enumerate]{noitemsep, topsep=0pt, parsep=0pt, partopsep=0pt}

\usetikzlibrary{decorations.pathreplacing}
\usetikzlibrary{shapes.geometric}
\usetikzlibrary{arrows}

\input{macro}
\newtheorem{theorem}{Theorem}
\newtheorem{lemma}{Lemma}

\theoremstyle{definition}
\newtheorem{definition}[theorem]{Definition}

\newtheorem{proposition}[lemma]{Proposition}

\newtheorem{example}{Example}[section]

\vldbTitle{Efficient Knowledge Graph Accuracy Evaluation}
\vldbAuthors{Junyang Gao, Xian Li, Yifan Ethan Xu, Bunyamin Sisman, Xin Luna Dong, Jun Yang}
\vldbDOI{https://www.doi.org/10.14778/3342263.3342642}
\vldbVolume{12}
\vldbNumber{11}
\vldbYear{2019}

\begin{document}

\title{Efficient Knowledge Graph Accuracy Evaluation}

\author{%
\alignauthor
Junyang Gao$^\dag$\titlenote{Most of the work was conducted when the author was interning at Amazon.}\hspace*{0.1em}
Xian Li$^\ddag$\hspace*{0.1em}
Yifan Ethan Xu$^\ddag$\hspace*{0.1em}
Bunyamin Sisman$^\ddag$\hspace*{0.1em}
Xin Luna Dong$^\ddag$\hspace*{0.1em}
Jun Yang$^\dag$\hspace*{0.1em}\\
$^\dag$Duke University, $^\ddag$Amazon.com\\
\vspace{1em}
\texttt{\{jygao,junyang\}@cs.duke.edu,}\hspace{1em}
\texttt{\{xianlee,xuyifa,bunyamis,lunadong\}@amazon.com}
}

\maketitle
\pagenumbering{gobble}

\begin{abstract}
Estimation of the accuracy of a large-scale knowledge graph (KG) often requires humans to annotate samples from the graph.
How to obtain statistically meaningful estimates for accuracy evaluation while keeping human annotation costs low is a problem critical to the development cycle of a KG and its practical applications.
Surprisingly, this challenging problem has largely been ignored in prior research.
To address the problem, this paper proposes an efficient sampling and evaluation framework,
which aims to provide quality accuracy evaluation with strong statistical guarantee while minimizing human efforts.
Motivated by the properties of the annotation cost function observed in practice, we propose the use of cluster sampling to reduce the overall cost.
We further apply weighted and two-stage sampling as well as stratification for better sampling designs.
We also extend our framework to enable efficient incremental evaluation on evolving KG, introducing two solutions based on stratified sampling and a weighted variant of reservoir sampling.
Extensive experiments on real-world datasets demonstrate the effectiveness and efficiency of our proposed solution.
Compared to baseline approaches,
our best solutions can provide up to 60\% cost reduction on static KG evaluation and up to 80\% cost reduction on evolving KG evaluation, without loss of evaluation quality.
\end{abstract}

\input{1-Introduction.tex}
\input{2-Preliminaries-v2.tex}
\input{3-Evaluation-cost-model-v2.tex}
\input{4-Evaluation-framework-v2.tex}
\input{5-Sampling-design-v2.tex}
\input{6-Incremental-evaluation-on-evolving-kg-v2.tex}
\input{7-Experiments-v2.tex}
\input{8-Related-work.tex}

\section{conclusion}\label{sec:conclusion}
In this paper, we have initiated a comprehensive study into the important problem of efficient and reliable knowledge graph accuracy evaluation.
We presented a general evaluation framework that works on both static and evolving KGs.
We devised a suite of sampling techniques for efficient accuracy evaluation on these two scenarios.
As demonstrated by experiments on various KGs with real and synthetic labels on triple correctness, our solutions can significantly speed up the accuracy evaluation process compared to existing baseline approaches.
\ansa{Future work includes extending the proposed solution to enable efficient evaluation on different granularity, such as accuracy per predicate or per entity type.}
\paragraph*{\textbf{Acknowledgements}}
Junyang Gao and Jun Yang were supported by NSF grants IIS-1408846, IIS-1718398, and IIS-1814493. Any opinions, findings, and conclusions or recommendations expressed in this publication are those of the author(s) and do not necessarily reflect the views of the funding agencies.
\newpage

\balance
\bibliographystyle{abbrv}
\bibliography{ref}  

\end{document}

%% file: macro.tex
\usepackage{euscript}
\usepackage{xparse}
\usepackage{mathrsfs}
\newcommand{\narrow}[1]{\protect\scalebox{0.7}[1.0]{\ensuremath{\textsf{#1}}}}
\newcommand{\card}[1]{\ensuremath{\lvert#1\rvert}}

\newcommand{\acc}{\mu}

\newcommand{\cost}{\narrow{Cost}}

\newcommand{\E}{\mathrm{E}}
\newcommand{\Var}{\mathrm{Var}}

\newcommand{\est}{\hat{\mu}}

\newcommand{\design}{\mathcal{D}}

\usepackage{amssymb}
\usepackage{pifont}
\newcommand{\cmark}{\ding{51}}%
\newcommand{\xmark}{\ding{55}}%

\definecolor{reva}{rgb}{0,0,0}
\definecolor{revb}{rgb}{0,0,0}
\definecolor{revc}{rgb}{0,0,0}
\definecolor{revmeta}{rgb}{0,0,0}
\definecolor{default}{rgb}{0,0,0}

\newcommand{\ansmeta}[1]{{{\color{revmeta}{#1}}}}
\newcommand{\ansa}[1]{{{\color{reva}{#1}}}}
\newcommand{\ansb}[1]{{{\color{revb}{#1}}}}
\newcommand{\ansc}[1]{{{\color{revc}{#1}}}}

%% file: 1-Introduction.tex
\section{Introduction}\label{sec:intro}

Over the past few years,
we have seen an increasing number of large-scale KGs with millions of relational facts  in  the  format of RDF triples \emph{(subject,predicate,object)}.
Examples include DBPedia~\cite{dbpedia}, YAGO~\cite{fabian2007yago,YAGO}, NELL~\cite{mitchell2018never}, Knowledge-Vault~\cite{dong2014knowledge}, etc.
However, the KG construction processes are far from perfect, 
so these KGs may contain many incorrect facts. 
Knowing the accuracy of the KG is crucial for improving its construction process (e.g., better understanding the ingested data quality and defects in various processing steps), and informing the downstream applications and helping them cope with any uncertainty in data quality. 
Despite its importance, the problem of efficiently and reliably evaluating KG accuracy has been largely ignored by prior academic research. 

KG accuracy can be defined as the percentage of triples in the KG being \emph{correct}. 
Here, we consider a triple being correct if the corresponding relationship is consistent with the real-life fact.
Typically, we rely on human judgments on the correctness of triples. 
Manual evaluation at the scale of modern KGs is prohibitively expensive. 
Therefore, the most common practice is to carry out manual annotations on a (relatively small) sample of KG and compute an estimation of KG accuracy based on the sample. 
A naive and popular approach is to randomly sample triples from the KG to annotate manually.
A small sample set translates to lower manual annotation costs, but it can potentially deviate from the real accuracy.
In order to obtain a statistically meaningful estimation, one has to sample a large ``enough'' number of triples, so increasing cost of annotation.
Another practical challenge is that KG evolves over time---as new facts are extracted and added to the KG, its accuracy changes accordingly.
Assuming we have already evaluated a previous version of the KG, we would like to incrementally evaluate the accuracy of the new KG without starting from scratch.

\begin{table*}
\centering
\setlength\tabcolsep{1pt}
\caption{Two annotation tasks: Task1 consists of triples regarding different entities while Task2 consists of triples about the same entity. \label{fig:motivating-example}}
\begin{tabular}{ |l|l| }
\hline
  Task1 & Task2 \\\hline
  (Michael Jordan, graduatedFrom, UNC) & (Michael Jordan, wasBornIn, LA) \\
  (Vanessa Williams, performedIn, Soul Food) & (Michael Jordan, birthDate, February 17, 1963) \\
  (Twilight, releaseDate, 2008) & (Michael Jordan, performedIn, Space Jam) \\
  (Friends, directedBy, Lewis Gilbert) & (Michael Jordan, graduatedFrom, UNC) \\
  (The Walking Dead, duration, 1h 6min) & (Michael Jordan, hasChild, Marcus Jordan) \\\hline
\end{tabular}
\vspace{-1em}
\end{table*}

To motivate our solution, let us examine in some detail how the manual annotation process works.  We use two annotation tasks shown in Table~\ref{fig:motivating-example} as examples.
\begin{example}\label{eg:annotation}
Mentions of real-life entities can be ambiguous.
For example, the first triple in Task1, the name ``Michael Jordan'' could refer to different people --- Michael Jordan the hall-of-fame basketball player or Michael Jordan the distinguished computer scientist?
The former was born in New York, while the latter was born in Los Angeles.
Before we verify the relationship between subject and object, the first task is to identify each entity.\footnote{In an actual annotation task, each triple is associated with some context information. Annotators need to spend time first identifying the subject, the object or both.}
If we assess a new triple on an entity that we have already identified, the total evaluation cost will be lower compared to assessing a new triple from unseen entities. 
For example, 
in Task2, all triples are about the same entity of Michael Jordan. 
Once we identify this Michael Jordan as the basketball player, annotators could easily evaluate correctness of these triples without further identifications on the subject. 
On the contrary, in Task1, five different triples are about five different entities. 
Each triple's annotation process is independent, and annotators need to spend extra efforts first identifying possible ambiguous entities for each of them, i.e., Friends the TV series or Friends the movie? Twilight the movie in 2008 or Twilight the movie in 1998? 
Apparently, given the same number of triples for annotations, Task2 takes less time. 
In addition, validating triples regarding the same entity would also be an easier task.
For example, a WiKi page about an actor/actress contains most of the person's information or an IMDb page about a movie lists its comprehensive features.
Annotators could verify a group of triples regarding the same entity all at once in a single (or limited number) source(s) instead of searching and navigating among multiple sources just to verify an individual fact. 

Hence, generally speaking, auditing on triples about the same entity (as Task2) can be of lower cost than on triples about different entities (as Task1).
Unfortunately,
given the million- or even billion-scale of the KG size, selecting individual triples is more likely to produce an evaluation task as Task1.
\end{example}

As motivated in the above example, when designing a sampling scheme for large KG, the number of sampled triples is no longer a good indicator of the annotation cost---instead, we should be mindful of the actual properties of the manual annotation cost function in our sampling design.  \textbf{Our contributions} are four-fold:
\begin{itemize}
\item We provide an iterative evaluation framework that is guaranteed to provide high-quality accuracy estimation with strong statistical consistency. 
Users can specify an error bound on the estimation result, and our framework iteratively samples and estimates. It stops as soon as the error of estimation is lower than user required threshold without oversampling and unnecessary manual evaluations.
\item Exploiting the properties of the annotation cost, we propose to apply cluster sampling with unequal probability theory that enables efficient manual evaluations. We quantitatively derive the optimal sampling unit size in KGs by associating it with approximate evaluation cost. 
\item The proposed evaluation framework and sampling technique can be extended to enable incremental evaluation over evolving KGs.
We introduce two efficient incremental evaluation solutions based on stratified sampling and a weighted variant of reservoir sampling respectively.
They both enable us to reuse evaluation results from previous evaluation processes, thus significantly improving the evaluation efficiency.
\item Extensive experiments on various real-life KGs, involving both ground-truth labels and synthetic labels, demonstrate the efficiency of our solution over existing baselines.
For evaluation tasks on static KG, our best solution cuts the annotation cost up to 60\%.
For evaluation tasks on evolving KG, incremental evaluation based on stratified sampling provides up to 80\% cost reduction.
\end{itemize}
To the best of our knowledge, this work is among the first to propose a practical evaluation framework that provides efficient, unbiased, and high-quality KG accuracy estimation for both static and evolving KGs.%
\ansc{
Though we mainly focus on accuracy evaluation of knowledge graphs, our proposed evaluation framework and sampling techniques are general and can be extended to relational databases (with appropriate notions of entities and relationships).
}

The rest of the paper is organized as follows.
Section~\ref{sec:prelim} reviews the key concepts of KG accuracy evaluation and formally defines the problem.
Section~\ref{sec:cost} proposes an evaluation model and analyzes human annotator's performances over different evaluation tasks that motivate our solution.%
\ansb{
Section~\ref{sec:overview} presents our general evaluation framework.
Section~\ref{sec:sampling} and Section~\ref{sec:evolving-kg} introduce a comprehensive suite of sampling techniques that lead to efficient quality evaluation on both static KG and evolving KG. 
}
Section~\ref{sec:expr} experimentally evaluates our solutions.
Finally, we review related work on KG accuracy evaluation in Section~\ref{sec:related-work} and conclude in Section~\ref{sec:conclusion}.

%% file: 2-Preliminaries-v2.tex
\section{Preliminaries}
\label{sec:prelim}

\subsection{Knowledge Graphs}
\label{sec:KGs}
We model knowledge graph $G$ as a set of \emph{triples} in the form of (\emph{subject}, \emph{predicate}, \emph{object}), denoted by $(s,p,o)$.
Formally, $G = \{t \mid t:(s,p,o)\}$. 
For example, in tuple {\small\textbf{(/m/02mjmr, /people/person\-/place\_of\_birth, /m/02hrh0\_)}},
{\small\textbf{/m/02mjmr}} is the Freebase id for Barack Obama, and
{\small\textbf{/m/02hrh0\_}} is the id for Honolulu.
Each entity in the KG is referred to unique id.
If the object of a triple is an entity, we call it a \emph{triple with entity property}.
On the contrary, a triple with an atomic object, such as a date, number, length, etc., is called a \emph{triple with data property}.
Next, let us define an \emph{entity cluster} as a set of triples with the same subject value $e$; i.e., $G[e]  = \{t \mid t:(s,p,o) \land s = e\}$. 
For a knowledge graph $G$ with $n$ distinct entities $E = \{e_1,e_2, \cdots, e_n\}$, we have $G = \bigcup_{e \in E} G[e]$.

A knowledge graph $G$ may evolve over time.
Changes to $G$ can be modeled using a (possibly infinite) sequence of triple-level updates.
In practice, updates often arrive in batches.
In this paper, we only consider triple insertions into $G$.\footnote{We add a new entity into KG by inserting new triples regarding the entity.}
Consider a batch $\Delta$ of triple-level insertions.
We cluster all insertions by their subject id such that each $\Delta_e$ only contains those insertions regarding the same subject id $e$, denoted as $\Delta_e= \{t\mid t:(s,p,o) \land s=e\}$. 
The evolved KG is represented as $G + \Delta = G \cup \bigcup_e \Delta_e$.

\subsection{KG Accuracy and Estimation}\label{sec:unbiased-ci}
The correctness of a triple $t \in G$ is denoted by a value function $f : t \to \{0,1\}$, where $1$ indicates correct and $0$ incorrect.
The KG accuracy is defined as the mean accuracy of triples $\acc(G):=\frac{1}{\lvert G \rvert}\sum_{t\in G} f(t).$


In this paper, we compute the value of $f(t)$ by manual annotation. 
However, 
it is infeasible to manually evaluate every triple to assess the accuracy of a large-scale KG. 
A common practice is to estimate $\mu(G)$ with an estimator $\est$ calculated over a relatively small sample $G' \subset G$, where $G'$ is drawn according to a certain sampling strategy $\mathcal{D}$. 
For instance, the simplest estimator is the mean accuracy of a simple random sample of the triples in $G$. 
For the purpose of evaluating the accuracy of $G$, we require $\est$ to be unbiased; that is, $E[\est] = \acc(G)$.
To quantify the uncertainties in the sampling procedure, a confidence interval (CI) should be provided for a single-valued point estimator.
There is no universal formula to construct CI for an arbitrary estimator. 
However, if a point estimator $\est$ takes the form of the mean of $n$ independent and identically distributed (i.i.d.) random variables with equal expectation $\mu$, then by the Central Limit Theorem,\footnote{To be more precise, the Central Limit Theorem is applicable when $n$ is large. 
A rule of thumb is $n > 30$. The size restriction can be relaxed when the distribution of a sample is approximately Gaussian. 
See a standard Statistics text book~\cite{casella2002statistical} for a formal definition.} an approximate $1 - \alpha$ CI of $\mu$ can be constructed as 
\begin{equation}
\label{eq:ci}
\est\pm z_{\alpha/2}\sqrt\frac{\sigma^2}{n},
\end{equation}
where $z_{\alpha/2}$ is the Normal critical value with right-tail probability $\alpha/2$, and $\sigma^2$ is the population variance. The half width of a CI is called the Margin of Error (MoE). 
In (\ref{eq:ci}), the MoE is $z_{ \alpha/2}\sqrt{\sigma^2/n}$.

\subsection{Problem Formulation}\label{sec:ps}
We now formally define the task of efficient KG accuracy evaluation.
Let $G' = \mathcal{D}(G)$ be a sample drawn using a sampling design $\design$, and $\est$ be an estimator of $\acc(G)$ based on $G'$.
Let $\cost(G')$ denote the manual cost of annotating the correctness of triples in $G'$. 
we are interested in the following problem:
\begin{definition}[Efficient KG Accuracy Evaluation]\label{def:efficient-kgeval}
Given a KG $G$ and an upper bound of MoE $\epsilon$ at confidence level $1-\alpha$,
\begin{align}
    &\underset{\design}{\text{minimize}}
    & & \E\bigg[\cost\big(\mathcal{D}(G)\big)\bigg] \label{eq:obj}\\
    &\text{subject to}
    & & \E[\est] = \acc(G), \text{ MoE}(\est,\alpha)\leq \epsilon. \notag
\end{align}
\end{definition}

For the case of evolving KG, suppose we have already evaluated $G$ using a sample $G'$, and since then $G$ has evolved to $G+\Delta$.
Our goal is to minimize the evaluation cost to estimate $\acc(G+\Delta)$ given that $\acc(G)$ has been estimated.
Let $\mathcal{D}(G + \Delta \mid G')$ be a sample drawn using a sampling design $\design$ given $G'$,
and $\est$ is the estimator of $\acc(G+\Delta)$ based on $\mathcal{D}(G + \Delta \mid G')$ (also given $\acc(G)$ estimated using $G'$).
The problem can be stated more precisely as follows.
\begin{definition}[Efficient Evolving KG Accuracy Evaluation]\label{def:efficient-kgeval-evolving}
\begin{align}
    &\underset{\design}{\text{minimize}}
    & & \E\bigg[\cost\bigg(\mathcal{D}\big((G + \Delta)\, \mid\, G'\big)\bigg)\bigg] \label{eq:obj-evolving}\\
    &\text{subject to}
    & & \E[\est] = \acc(G+\Delta),\text{ MoE}(\est,\alpha)\leq \epsilon. \notag
\end{align}
\end{definition}

%% file: 3-Evaluation-cost-model-v2.tex
\section{Evaluation Cost Model}
\label{sec:cost}

Prior research typically ignores the evaluation time needed by manual annotations.
In this section, we study human annotators' performance on different evaluation tasks and propose a cost function that approximates the manual annotation time.
Analytically and empirically, we argue that
annotating triples in groups of entities is more efficient than triple-level annotation. 

\subsection{Evaluation Model}\label{sec:eval-model}
\begin{figure}
    \centering
    \includegraphics[width=0.45\textwidth]{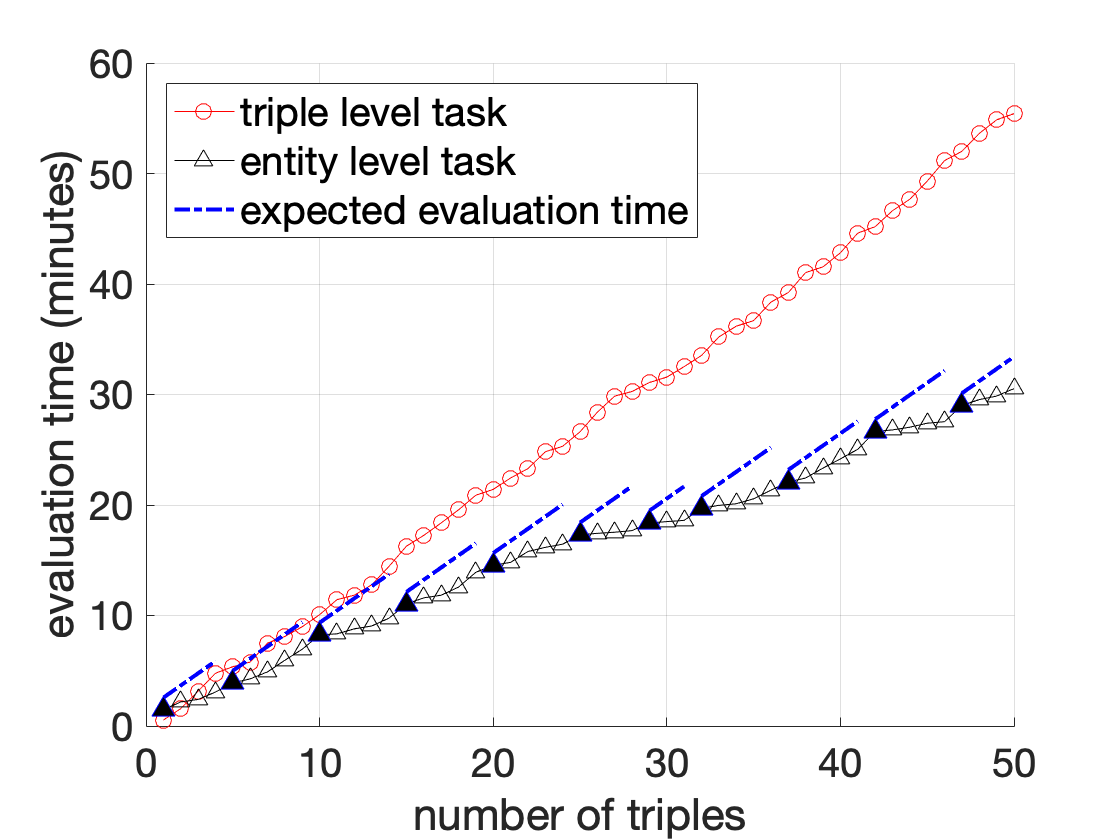}   
    \vspace{-1em}
    \caption{Evaluation cost comparison of triple-level and entity-level tasks on MOVIE. For entity-level tasks, the first triple evaluated from an entity cluster is marked as solid triangle.}
    \label{fig:eval-cost}
    \vspace{-0.5em}
\end{figure}
Recall that subject or non-atomic object in the KG is represented by an id, which refers to a unique real-life entity.
When manually annotating a $(s,p,o)$ triple, a connection between the id and the entity to which it refers must be first established.
We name this process as \emph{Entity Identification}.
The next step is to collect evidence and verify the facts stated by the triple,
which is referred to as \emph{Relationship Validation}.
\ansa{
To exploit the property of annotation cost as we motivated in Example~\ref{eg:annotation}, sampled triples are prepared by their subjects for manual evaluations. 
We shall refer to the task of manually annotating true/false labels for a group of triples with the same subject id as an \emph{Evaluation Task}.
}
In this paper, we consider the following general evaluation instructions for human annotators:
\begin{itemize}
    \item Entity Identification:
    Besides assigning annotators an Evaluation Task to audit, 
    we also provide a small set of related information regarding to the subject of this Task.
    Annotators are required to use the provided information to construct a clear one-to-one connection between the subject and an entity using their best judgement, 
    especially when there is ambiguity; that is, different entities share the same name or some attributes.
    \item Relationship Validation:
    This step asks annotators for a cross-source verification; that is, searching for evidence of subject-object relationship from multiple sources (if possible) and making sure the information regarding the fact is correct and consistent.
    Once we have a clear context on the Evaluation Task from the first step of Entity Identification, relationship validation would be a straightforward yes or no judgement.
\end{itemize}
\begin{example}\label{eg:manual-annotation}
We ask one human annotator to perform several annotation tasks on the MOVIE KG,\footnote{MOVIE is a knowledge graph constructed from IMDb and WiKiData. More detailed information can be found in Section~\ref{sec:expr:dataset}.} and track the cumulative time spent after annotating each triple.
In the first task (which we call ``triple level''), we draw 50 triples randomly from the KG, and ensure that all have distinct subject ids. 
In the second task (which we call ``entity level''), we select entity clusters at random, and from each selected cluster draw at most 5 triples at random; the total number of triples is still 50, but they come from only 11 entity clusters.
The cumulative evaluation time is reported in Figure~\ref{fig:eval-cost}.

The time required by evaluating triple-level task increases approximately linearly in the number of triples,
and is significantly longer than the time required for entity-level task, as we expected.
If we take a closer look at the plot for the entity-level task, it is not difficult to see that the evaluation cost on subsequent triples from an identified entity cluster is much lower on average compared to independently evaluating a triple (straight dotted lines).
\end{example}

\subsection{Cost Function}\label{sec:cost-func}
We define a cost function based on the proposed evaluation model. 
\begin{definition}[Evaluation Cost Function]
Given a sampled subset $G'$ from KG, the approximate evaluation cost is defined as
\begin{equation}\label{eq:simplified-cost}
    \cost(G') = \card{E'}\cdot c_1 + \card{G'}\cdot c_2,
\end{equation}
where $E'$ is the set of distinct ids from $G'$. $c_1, c_2$ are the average cost of entity identification and relationship validation, respectively.
\end{definition}
\vspace{-0.5em}
Average costs $c_1$ and $c_2$ are calculated from empirical evaluation costs by human annotators over multiple evaluation tasks.
In reality, the cost of evaluating triples vary by different human annotators, but in practice we found taking averages is adequate for the purpose of optimization because they still capture the essential characteristics of the cost function.
More details on the annotation cost study can be found in the experiment section.

%% file: 4-Evaluation-framework-v2.tex
\section{Evaluation Framework}\label{sec:overview}
\begin{figure}
\includegraphics[width=0.5\textwidth]{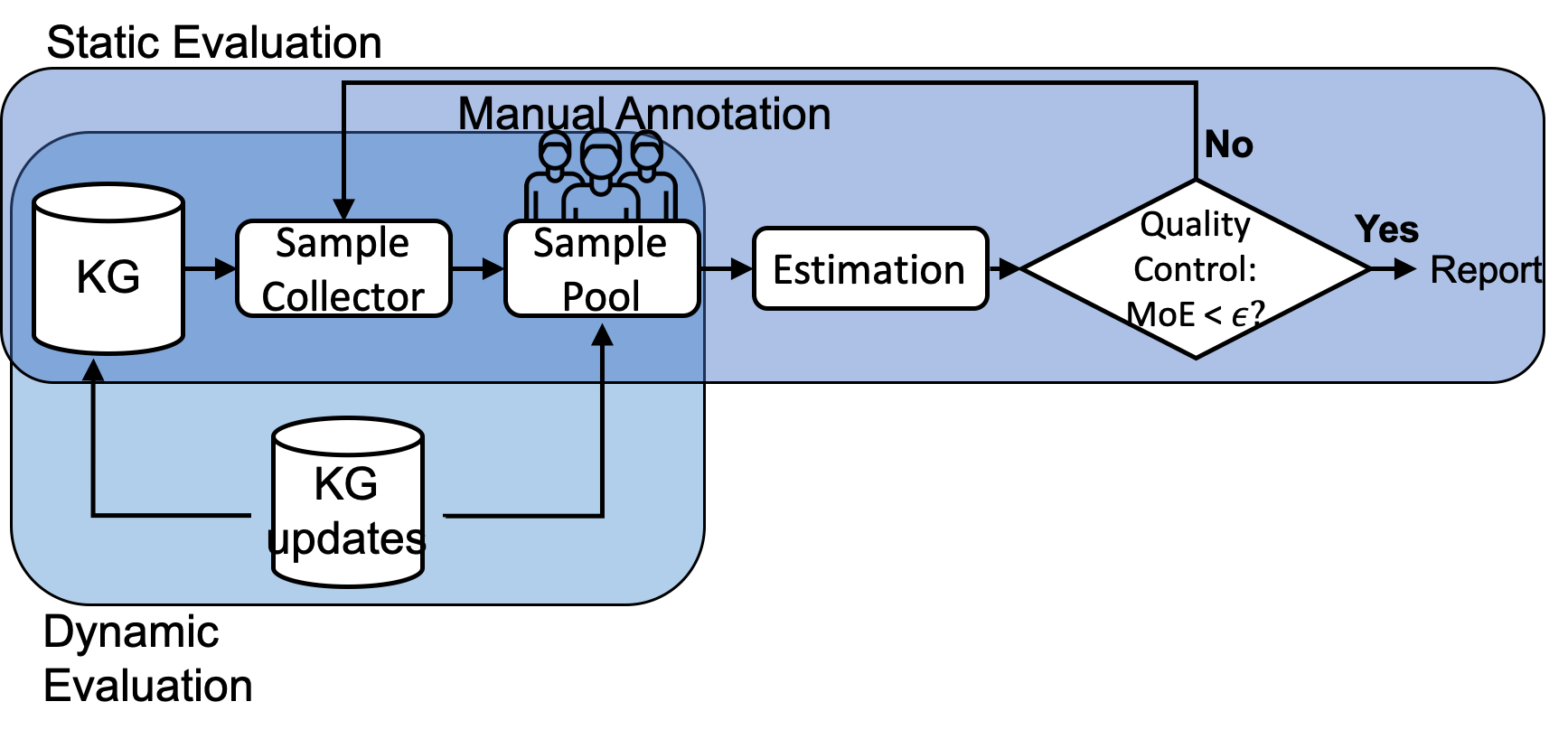}
\vspace{-1em}
\caption{Iterative KG accuracy evaluation framework.\label{fig:framework}}
\vspace{-1em}
\end{figure}

In a nutshell, our evaluation framework is shown in Fig~\ref{fig:framework}. There are two evaluation procedures.

\textbf{Static Evaluation} conducts efficient and high-quality accuracy evaluation on static KGs.
It works as the following iterative procedure, sharing the same spirit as in Online Aggregation~\cite{hellerstein1997online}:
\begin{itemize}
\item Step 1: \textbf{Sample Collector} selects a small batch of samples from KG using a specific sampling design $\mathcal{D}$.
Section~\ref{sec:sampling} discusses and compares various sampling designs.
\item Step 2: \textbf{Sample Pool} contains all samples drawn so far and asks for manual annotations when new samples are available.
\item Step 3: Given accumulated human annotations and the sampling design $\mathcal{D}$, the \textbf{Estimation} component computes an unbiased estimation of KG accuracy and its associated MoE.
\item Step 4: \textbf{Quality Control} checks whether the evaluation result satisfies the user required confidence interval and MoE. If yes, stop and report; otherwise, loop back to Step 1.
\end{itemize}

\textbf{Dynamic Evaluation} enables efficient incremental evaluation on evolving KGs. 
We propose two evaluation procedures. One is based on reservoir sampling, and the other is based on stratified sampling. Section~\ref{sec:evolving-kg} introduces the detailed implementations. 

\ansa{
It is worth mentioning that the proposed framework is generic and independent of the manual annotation process. 
Users can specify either single evaluation or multiple evaluations (assigned to different annotators) per Evaluation Task, as long as correctness labels for sampled triples are collected after manual annotations.
}

The proposed framework has the following advantages. (1) The framework selects and estimates iteratively through a sequence of small batches of samples from KG,
and stops as soon as the estimation quality satisfies the threshold (specified by MoE) as user required.
It avoids oversampling and unnecessary manual annotations and always reports an accuracy evaluation with strong statistical guarantee.
(2) The framework works efficiently both on static KGs and evolving KGs with append-only changes. 

%% file: 5-Sampling-design-v2.tex
\section{Sampling Design}\label{sec:sampling}
\begin{table}
    \centering
    \caption{Notations.}
    \label{tab:notation}
    \begin{tabular}{|r|c|}
    \hline
        $N$ &  Number of entity clusters in $G$\\\hline
        $M_i$ & Size of the $i$th entity cluster \\\hline
        $M = \sum_{i=1}^{N}M_i$ & Total number of triples in $G$ \\\hline
        $n$ & \makecell{Number of entity clusters \\in the sample}\\\hline
        $m$ & \makecell{Maximum number of triples \\to draw within each cluster} \\\hline
        $\tau_i$ & \makecell{Number of correct triples  \\in the $i$th cluster} \\\hline
        $\mu_i = \tau_i/M_i$ & Accuracy of the $i$th cluster \\\hline
    \end{tabular}
    \vspace{-1em}
\end{table}
The sampling design is the core component of the framework. 
There are two sampling strategies to evaluate the quality of KGs: triple-level and entity-level.
\textbf{Simple random sampling} (SRS) is the most common triple-level sampling design, while \textbf{Cluster sampling} (CS) samples entity clusters instead of triples. In this section we present one SRS-based estimator and three CS-based estimators.
Frequently used notations are summarized in Table~\ref{tab:notation}.

\subsection{Simple Random Sampling}\label{sec:srs}
With SRS, we randomly draw $n_s$ triples $t_1\ldots,t_{n_s}$ from $G$ without replacement. An unbiased estimator of $\acc(G)$ is the sample mean
\begin{equation}
\est_s:=\frac{1}{n_s}\sum_{i=1}^{n_s}f(t_i).
\end{equation}
With \emph{Normal approximation}, its $1-\alpha$ CI is 
$
    \est_s \pm z_{\alpha/2} \sqrt{\frac{\est_s(1-\est_s)}{n_s}}.
$
\paragraph*{Cost Analysis}
Note that even though we draw each triple independently, in practice when human annotators carry out the task, we still group triples by subject ids to save the entity identification costs. The expected number of unique entities in a SRS sample is
\[
    \E[n_c] = \sum_{i=1}^N\bigg(1 - \left(1 - \frac{M_i}{M}\right)^{n_s}\bigg).
\]
Thus the objective in (\ref{eq:obj}) can be rewritten as
\begin{align}
    &\underset{n_s}{\text{minimize}}
    & & \sum_{i=1}^N\bigg(1 - \left(1 - \frac{M_i}{M}\right)^{n_s}\bigg)c_1 + n_s c_2 \label{eq:srs-obj-full}\\
    &\text{subject to}
    & & \text{MoE}(\est_s,\alpha) \leq \epsilon. \notag
\end{align}
Since the objective in (\ref{eq:srs-obj-full}) is monotonically increasing in $n_s$,
the minimum is achieved at 
\[
n_s = \frac{\est_s(1 - \est_s)z^2_{\alpha/2}}{\epsilon^2}.
\]
A drawback of SRS is that triples are randomly scattered among entities, requiring a large number of entities to be identified hence incurring higher identification cost. We can reduce this cost by intentionally sampling more triples from the same entity clusters.

\subsection{Cluster Sampling}\label{sec:cs}
In general, auditing triples from the same entity costs less than auditing the same amount of triples from different entities, as shown in Example~\ref{eg:annotation} and in Figure~\ref{fig:eval-cost}.
For this reason, it is natural to derive estimators based on cluster sampling (CS) schemes, where triples from the same entity are drawn together. 
We introduce estimators based on three cluster sampling strategies, \emph{random cluster sampling}, \emph{weighted cluster sampling}, and \emph{two-stage cluster sampling}, and reason that the last has the lowest cost. 

\subsubsection{Random Cluster Sampling}\label{sec:rcs}
With random cluster sampling (RCS), $n$ entity clusters are drawn randomly, and all triples in the sampled clusters are manually evaluated. 
Let $I_k$ be the index of the $k$-th sample cluster, $k=1,2,\dots,n$.
An unbiased estimator of $\acc(G)$ is given by
\begin{equation}\label{eq:rcs-unbiased}
\small
    \est_r := \frac{N}{Mn}\sum_{k=1}^{n} \tau_{I_k}
\end{equation}
and its $1-\alpha$ CI is
$
\est_r \pm z_{\alpha/2}\sqrt{\frac{1}{n(n-1)}\sum_{k=1}^{n}(\frac{N}{M}\tau_{I_k} - \est_r)^2}.
$

Since $\est_r$ relies on the number of correct triples $\tau_I$ in each sampled cluster, which is positively correlated to the cluster size, the variance of $\est_r$ is high when the cluster size distribution is wide, 
which is the case of most real-life KGs.
A more robust estimator in such situations is based on the proportion rather than the number of correct triples in sampled clusters. 

\subsubsection{Weighted Cluster Sampling}\label{sec:wcs}
With weighted clustering sampling (WCS), clusters are drawn with probabilities proportional to their sizes: $\pi_i=M_i / M, i=1, ..., N$. Then all triples in sampled clusters are evaluated. An unbiased estimator of $\acc(G)$ is the \emph{Hansen-Hurwitz estimator}~\cite{hansen1943theory}:
\begin{equation}\label{eq:wcs-unbiased}
\small
    \est_w := \frac{1}{n}\sum_{k=1}^{n} \mu_{I_k}.
\end{equation}
The $1-\alpha$ CI of $\est_w$ is
$
    \est_w \pm z_{\alpha/2} \sqrt{\frac{1}{n(n-1)}\sum_{k=1}^{n} (\mu_{I_k} - \est_w)^2}.
$

Comparing to \eqref{eq:rcs-unbiased}, $\est_w$ has a smaller variance when cluster sizes have a wide spread, because $\est_w$ sums over the accuracies of clusters rather than the number of accurate triples of sampled clusters.  

\subsubsection{Two-Stage Weighted Cluster Sampling}\label{sec:2-wcs}

The cost of WCS can be further reduced by estimating the accuracies of sampled clusters from samples of triples, instead of evaluating every single triple in the cluster. 
The cost saving from the second stage within-cluster sampling is especially significant when KG contains large entity clusters with hundreds or even thousands of triples, which is common in most KGs. 
In this section, we introduce two-stage weighted cluster sampling (TWCS):
\begin{enumerate}
    \item In the first stage, we sample entity clusters using WCS.
    \item In the second stage, only a small number of triples are selected randomly from clusters sampled in the first stage.  More specifically, $\min\{M_{I_k},m\}$ triples are randomly drawn from the $k$-th sample cluster \emph{without replacement}.
\end{enumerate}
Drawing without replacement in the second stage greatly reduces sampling variances when cluster sizes are comparable or smaller than $m$. 
The finite population correction factor is applied to subsequent derivations accordingly.
(A similar approach can be applied to two-stage random cluster sampling; however, due to its inferior performance, we omit the discussion.)

Next, we show that TWCS still provides an unbiased estimation.
Let $\hat\mu_{I_k}$ be the mean accuracy of the sampled triples (at most $m$) in the k-th sampled cluster. An unbiased estimator of $\acc(G)$ is
\begin{equation}\label{eq:twcs-estimator}
    \est_{w,m} = \frac{1}{n}\sum_{k=1}^{n}\hat{\mu}_{I_k}.
\end{equation}
The $1-\alpha$ CI of $\est_{w,m}$ is
\[
    \est_{w,m} \pm z_{\alpha/2} \sqrt{\frac{1}{n(n-1)}\sum_{k=1}^{n} (\hat{\mu}_{I_k} - \est_{w,m})^2}.
\]
\begin{proposition}
Using TWCS with a second-stage sample size of $m$, $\est_{w,m}$ is an unbiased estimator of $\acc(G)$; that is, $\E[\est_{w,m}] = \acc(G)$.
\end{proposition}
\begin{proof}
By linearity of expectation, $\E[\est_{w,m}] = \frac{1}{n}\sum_{k=1}^{n}\E[\hat{\mu}_{I_k}]$.
Since SRS is applied in each selected cluster, $\E[\hat{\mu}_{I_k}] = \E[\mu_{I_k}]$.
Finally, each cluster is sampled with probability of $\frac{M_i}{M}$,
\[
\E[\est_{w,m}] = \frac{1}{n}\sum_{k=1}^{n}\E[\mu_{I_k}] = \frac{1}{n}\sum_{k=1}^{n} \sum_{i=1}^{N} \frac{M_i}{M} \mu_i
= \frac{1}{n}\sum_{k=1}^{n} \acc(G) = \acc(G).
\]
\end{proof}
\vspace{-1em}
The variance derivation of $\est_{w,m}$ is non-trivial.
Because of space limits, we move the full derivations into the extended version of this paper~\cite{fullversion}.
The theoretical variance of $\est_{w,m}$ is
\begin{equation}\label{eq:wcs-var-final}\resizebox{0.9\hsize}{!}{$%
     \Var(\est_{w,m})
     = \frac{1}{nM}\bigg(\sum_{i=1}^{N}M_i(\mu_i - \mu)^2 + \frac{1}{m}\sum_{i:M_i > m} \frac{M_i-m}{M_i-1}\cdot M_i \cdot \mu_i(1-\mu_i)\bigg)%
$}\end{equation}
\vspace{-1em}
\paragraph*{Relationship to SRS}
There is a close connection between SRS and TWCS, as the following result shows. Due to space limits, see the extended version of this paper~\cite{fullversion} for proof.
\begin{proposition}\label{lemma:wcs-to-srs}
The two-stage weighted cluster sampling with $m=1$ is equivalent to simple random sampling.
\end{proposition}
\vspace{-1em}
\paragraph*{Cost Analysis}
Now we derive the manual annotation cost required by TWCS.
Due to the second-stage sampling procedure, explicitly writing the cost function under TWCS would unnecessarily complicate the optimization objective.
Instead, we minimize an upper bound (achieved when all sample clusters contains at least $m$ triples) on the exact cost.
The objective in (\ref{eq:obj}) can be rewritten as:
\begin{align}
    &\underset{n,m}{\text{minimize}}
    & & n c_1 + (nm) c_2 \label{eq:wcs-obj-upper}\\
    &\text{subject to}
    & & \text{MoE}(\est_{w,m},\alpha) \leq \epsilon. \notag
\end{align}
Recall that the variables $n$ and $m$ are constrained by $\Var(\est_{w,m})$ in (\ref{eq:wcs-var-final}).
Thus the constraint in the optimization problem on MoE can be used to express $n$ as a function of $m$,
\[
n \geq \frac{V(m)z^2_{\alpha/2}}{\epsilon^2},
\]
where
\[
    V(m) = \frac{1}{M}\bigg(\sum_{i=1}^{N}M_i(\mu_i - \mu)^2 + \frac{1}{m}\sum_{i:M_i > m} \frac{M_i-m}{M_i-1}\cdot M_i \cdot \mu_i(1-\mu_i)\bigg).
\]
Since the minimization objective~(\ref{eq:wcs-obj-upper}) monotonically increases with $n$, 
it is sufficient to set
$n = V(m)z^2_{\alpha/2}/\epsilon^2$.
Hence, the objective can be further rewritten as
\begin{align}
    &\underset{m}{\text{minimize}}
 & \frac{V(m)z^2_{\alpha/2}}{\epsilon^2}(c_1 +  m c_2). \label{eq:wcs-obj-upper-m}
\end{align}
Though it is hard to provide a closed-form expression for the optimal $m$, we can easily find it via gradient descent or linear search on the discrete variable space.
In Section~\ref{sec:expr}, our experimental results confirm that optimizing the upper bound of the approximate cost can find $m$ very close to the empirical optimal solution.

Compared with the evaluation cost required by SRS as in~(\ref{eq:srs-obj-full}), the main cost saving provided by TWCS is on the triple identification cost specified as $c_1$. 
In order to achieve the MoE of estimation required by users, $n$ by TWCS is significantly smaller than $\sum_{i=1}^N \smash{\left(1 - (1 - \frac{M_i}{M})^{n_s} \right)}$.
Even though TWCS might eventually annotate more triples than SRS (because we have to annotate up to $m$ triples for each sample cluster), TWCS still beats SRS in terms of overall cost as auditing triples related to the same entity is efficient for human annotators.

\subsection{Further Optimization: Stratification}\label{sec:stratified}
\begin{figure}
\centering
\subfloat[NELL]{\includegraphics[width=0.25\textwidth]{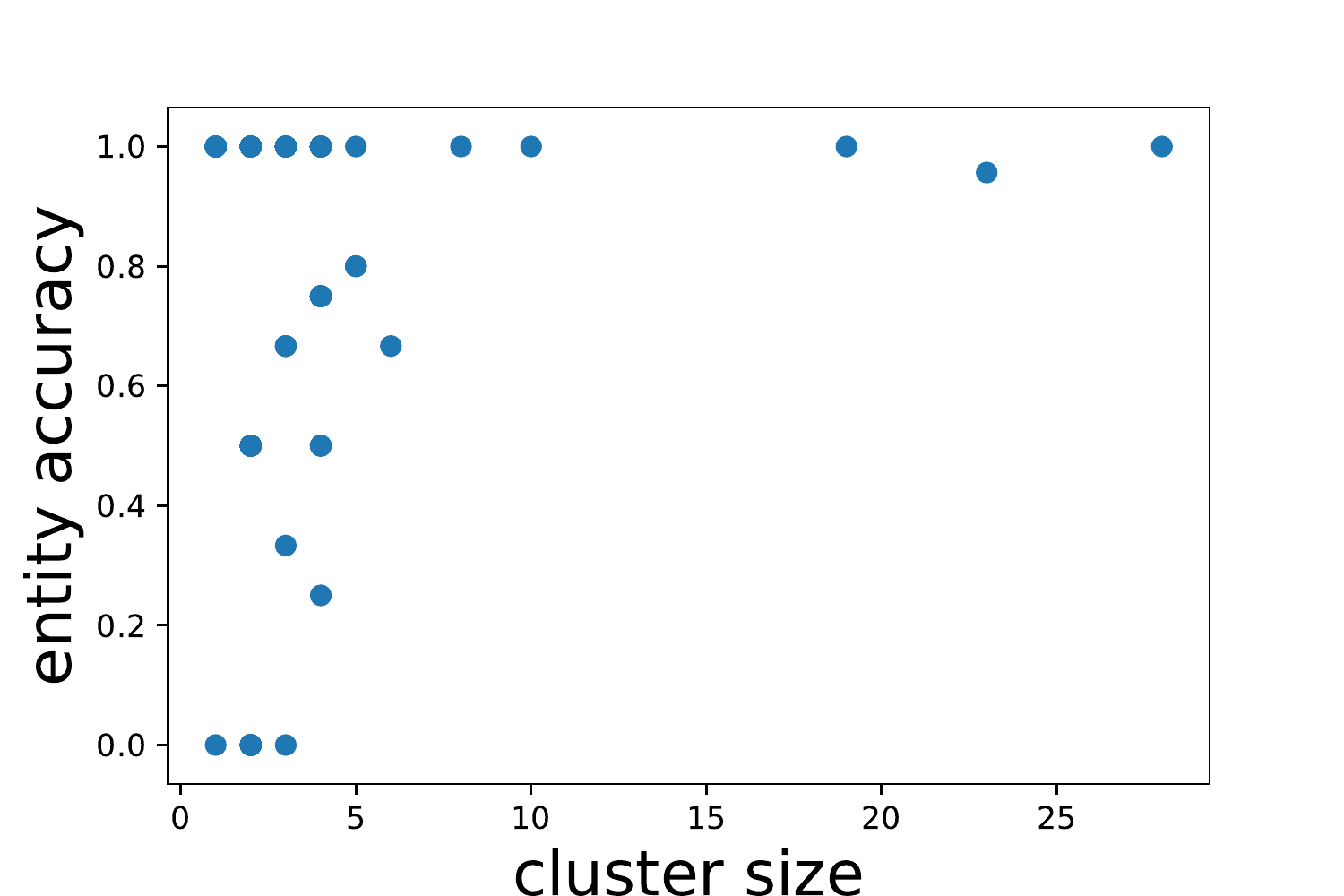}}
\subfloat[YAGO]{\includegraphics[width=0.25\textwidth]{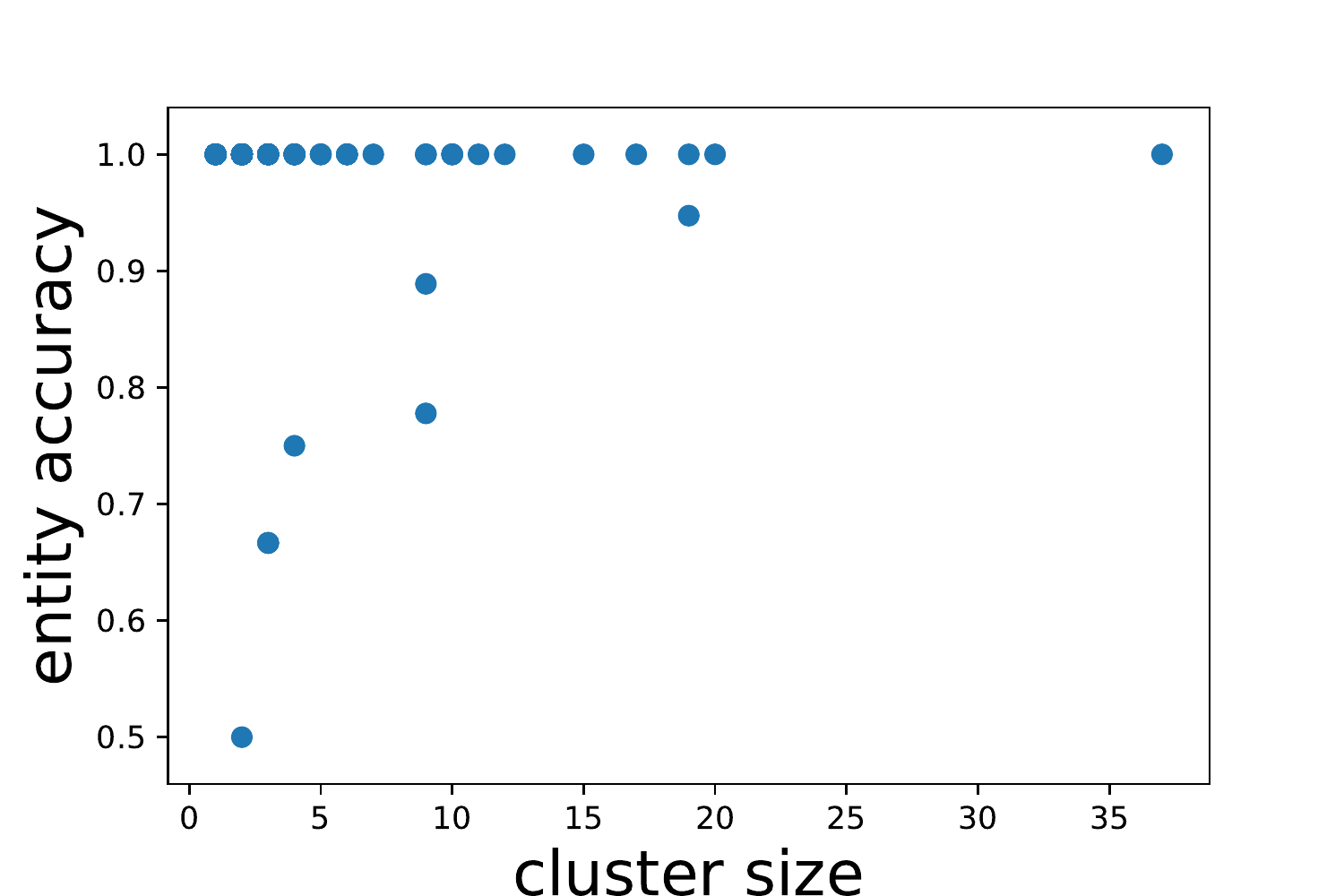}}
\vspace{-1em}
\caption{Correlation between entity accuracy and cluster size in real-life KGs, NELL and YAGO. Entity accuracy is defined as the percentage of triples being correct in the entity cluster. \label{fig:correlation}}
\vspace{-1.5em}
\end{figure}
In Figure~\ref{fig:correlation}, 
we plot the relationship between entity cluster accuracy and cluster size based on two real-life knowledge graphs NELL and YAGO with human annotated labels on triple correctness. 
We observe that larger entity clusters tend to have higher entity accuracy and also lower variation on entity accuracies. 
Motivated by this example,
cluster size seems to be a reasonable signal for gauging entity accuracies; e.g., large clusters tend to be more accurate. Such a signal would guide us to group clusters into sub-populations, within which entity accuracies are relatively homogeneous compared to the overall population. \emph{Stratified sampling}, a technique of variance reduction in statistics~\cite{casella2002statistical}, can then be applied to further boost the efficiency of cluster sampling.

Suppose we stratify entity clusters in KG into $H$ non-overlapping \emph{strata}. 
In each stratum $h$, we apply TWCS with second-stage sampling size $m$ to obtain an unbiased estimator $\est_{w,m,h}$.
Let $M[h]$ be the total number of triples in the $h$-th stratum, $W_h = M[h]/M$ as the weight of $h$-th stratum,
an unbiased estimator for KG accuracy is:
\begin{equation}\label{eq:ss-estimator}
    \est_{ss} = \sum_h W_h \cdot \est_{w,m,h},
\end{equation}
and its $1-\alpha$ CI is 
$
    \est_{ss} \pm z_{\alpha/2} \sqrt{\sum_h W_h^2 \cdot \Var(\est_{w,m,h})}.
$

If these strata are relatively homogeneous compared to the entire population, i.e., entity cluster with similar accuracy are grouped into the same stratum, then $\sum_h  W_h^2 \Var(\est_{w,m,h}) < \Var(\est_{w,m})$.
Thus, to achieve similar variance as in objective (\ref{eq:wcs-obj-upper}), we can get a even lower sample size $n$ to further reduce the annotation cost.

The interested readers can refer to the extended version of this paper~\cite{fullversion} for additional discussions on how to apply stratified sampling and optimal sample allocation.

Besides the cluster size, other signals (such as entity type, popularity, freshness) may also be useful for predicting accuracy and guiding the stratification strategy. Applying stratified sampling to SRS is also possible, although it would require a triple-level accuracy model, which is more difficult. How to design better accuracy models to better guide stratification is a promising direction of future work, but is beyond the scope of this paper.

%% file: 6-Incremental-evaluation-on-evolving-kg-v2.tex
\section{Evaluation on Evolving KG}\label{sec:evolving-kg}
We discuss two methods to reduce the annotation cost in estimating  $\acc(G+\Delta)$ as a KG evolves from $G$ to $G+\Delta$, one based on reservoir sampling and the other based on stratified sampling. 

\subsection{Reservoir Incremental Evaluation}\label{sec:reservoir-incremental}
Reservoir Incremental Evaluation (RS) is based on the \emph{reservoir sampling} \cite{vitter1985random} scheme, which stochastically updates samples in a fixed size reservoir as the population grows. 
To apply reservoir sampling on evolving KG to obtain a sample for TWCS, 
We introduce the reservoir incremental evalution procedure as follows.

For any batch of insertions $\Delta_e \in \Delta$ to an entity $e$, we treat $\Delta_e$ as a new and independent cluster, despite the fact that $e$ may already exist in the KG. This is to ensure weights of clusters stay constant. 
Though we may break an entity cluster into several disjoint sub-clusters over time, it does not change the properties of weighted reservoir sampling or TWCS, since these sampling techniques work independently on the definition of clusters. 

The Reservoir Incremental Evaluation procedure based on TWCS reservoir sampling is described in Algorithm~\ref{algo:reservoir}.
\vspace{-1em}
\begin{algorithm}
 \KwIn{\\A base knowledge graph $G$,\\ A TWCS sampled cluster set $R=\{r_1,\ldots,r_n\}$ with reservoir key value $K = \{k_1, \ldots, k_n\}$ generated by Algorithm-A in~\cite{efraimidis2006weighted},\\ A KG update $\Delta$.}
 \KwOut{A weighted random entity cluster sample $R$ of $G+\Delta$.}

 \For{$\Delta_e \in \Delta$} {
    $G \leftarrow G \cup \Delta_e$;
    
    Find the smallest reservoir key value in $K$ as $k_j$;
    
    Compute update key value $k_e=[\mathrm{Rand}(0, 1)]^{1/|\Delta_e|}$ ;
    
    \If{$k_e > k_j$}{
        $r_j\leftarrow \Delta_e$;
        
        $k_j\leftarrow k_e$;
    }
 }
 \KwRet $R$\;
 \caption{Reservoir-based Incremental Sample Update on Evolving KG\label{algo:reservoir}}
\end{algorithm}
\vspace{-1em}

Two properties of RS make it preferable for dynamic evaluation on evolving KGs.
First, as $G$ evolves, RS allows an efficient one-pass scan over the insertion sequence to generate samples.
Secondly, compared to re-sampling over $G+\Delta$, RS retains a large portion of annotated triples in the new sample, thus significantly reduces annotation costs. 

After incremental sample update using Algorithm~\ref{algo:reservoir}, it happens that the MoE of estimation becomes larger than the required threshold $\epsilon$.
In this case, 
we again run Static Evaluation process on $G+\Delta$ to draw more batches of cluster samples from the the current state of KG iteratively until MoE is no more than $\epsilon$.

\paragraph*{Cost Analysis}
As KG evolves from $G$ to $G + \Delta$, Algorithm~\ref{algo:reservoir} incrementally updates the random sample $G'$ to $(G+\Delta)'$, which avoids a fresh round of manual annotation.
Our estimation process only needs an incremental evaluation on these (potentially small) newly sampled entities/triples. 
Also in \cite{efraimidis2006weighted}, the authors pointed out that the expected number of insertions into the reservoir (without the initial insertions into an empty reservoir) is
\begin{align}
\begin{split}
    \text{\# of insertions} &= \sum_{i=N_i}^{N_j} \Pr[\text{cluster $i$ is inserted into reservoir}] \\
    &= O\bigg(\card{R} \cdot \log{\bigg(\frac{N_j}{N_i}}\bigg)\bigg),
\end{split}
\end{align}
where $\card{R}$ is the size of reservoir, and $N_i,N_j$ is the total number of clusters in $G, G+\Delta$, respectively.

\begin{proposition}\label{lemma:incremental-cost-bound}
    The incremental evaluations on new samples incurred by weighted sample update on evolving KG is at most \\
    $O\bigg(\card{R} \cdot \log{\big(\frac{N_j}{N_i}}\big)\bigg)$, where $R$ is the origin sample pool, and $N_i, N_j$ is the total number of clusters in the origin KG and the evolved KG, respectively.
    \vspace{-0.5em}
\end{proposition}

\ansb{
\subsection{Stratified Incremental Evaluation}\label{sec:stratified-incremental}
We now consider an even more efficient incremental evaluation method based on stratified sampling.
KG updates come in batches, and it is natural to view each batch of updates as a stratum. 
More specifically, when $G$ evolves to $G+\Delta$, $G$ and $\Delta$ are two independent and non-overlapping strata.
Stratified sampling enables us to combine the estimation results from $G$ and $\Delta$ to calculate the unbiased estimation of overall accuracy of $G+\Delta$.
Suppose that from previous round of evaluation on $G$, we have already collected a set of samples, and calculated $\hat{\mu}(G)$ and $\Var[\hat{\mu}(G)]$.
To evaluate overall accuracy of $G+\Delta$, we can fully reuse the samples drawn from $G$ (or more precisely, $\hat{\mu}(G)$ and $\Var[\hat{\mu}(G)]$), and only sample and annotate a few more triples on $\Delta$.
Guaranteed by stratified sampling, we can still have an unbiased estimation of the overall accuracy.
The full description of Stratified Incremental Evaluation procedure is shown as Algorithm~\ref{algo:stratified}.
\vspace{-1em}
\begin{algorithm}
\ansb{
 \KwIn{A base knowledge graph $G$, a batch of KG update $\Delta$, user-required MoE threshold $\epsilon$.}
 \KwOut{An unbiased estimation $\hat{\mu}(G+\Delta)$ of overall accuracy of $G+\Delta$, with MoE$\leq \epsilon$.}
 
 From accuracy evaluation on $G$, get $\hat{\mu}(G)$ and $\Var[\hat{\mu}(G)]$\;
 Calculate strata weight: $w_G = \frac{\card{G}}{\card{G+\Delta}}$, $w_\Delta = \frac{\card{\Delta}}{\card{G+\Delta}}$\;
 Initialization: $\hat{\mu}(G+\Delta)\leftarrow 0, MoE\leftarrow 1, S_\Delta \leftarrow \emptyset$ \;
 \While {$MoE > \epsilon$} {
    Randomly draw a batch of samples $B_\Delta$ using TWCS on $\Delta$ and append to sample set $S_\Delta$ : \newline
    $S_\Delta \leftarrow S_\Delta \cup B_\Delta$ \;
    Calculate $\hat{\mu}(\Delta)$ and $\Var[\hat{\mu}(\Delta)]$ from $S_\Delta$ using Eq(\ref{eq:twcs-estimator})\;
    Update $\hat{\mu}(G+\Delta)$ and $MoE$ using Eq(\ref{eq:ss-estimator})\;
 }
 \KwRet $\hat{\mu}(G+\Delta)$, $MoE$\;
 \caption{Stratified Incremental Evaluation on Evolving KG\label{algo:stratified}}
 }
\end{algorithm}
\vspace{-1em}

Though Algorithm~\ref{algo:stratified} only shows how to handle a single update batch, it can be extended straightforwardly to handle a sequence of KG updates over time.
Suppose we need to monitor the overall accuracy of evolving KG over a sequence of updates: $\Delta^1, \Delta^2, ..., \Delta^n$.
Each $\Delta^i$ will be treated as an independent stratum for stratified sampling evaluation.
For example, after applying $\Delta^i$, there is a total number of $i+1$ strata: $\{G, \Delta^1,..., \Delta^i\}$.
Similarly as in Algorithm~\ref{algo:stratified}, we reuse the evaluation results from strata $G,\Delta^1,...,\Delta^{i-1}$ and only incrementally draw samples from $\Delta^i$.

Compared with RS, Stratified Incremental Evaluation (SS) fully leverages annotations and evaluation results from previous rounds, without discarding any annotated triples.
That is the main reason why SS can be more efficient than RS.
Our experiments in Section~\ref{sec:expr} suggests that SS can bring a 20\% to 67\% improvement in evaluation efficiency compared to RS.
On the other hand, precisely because SS reuses \emph{all} previously annotated triples, it is more susceptible to the problem that a subset of samples may have a long-term impact on the quality of subsequent estimations. This trade-off between SS and RS is further evaluated in Section~\ref{sec:expr}.
}

%% file: 7-Experiments-v2.tex
\section{Experiments}\label{sec:expr}
In this section, we comprehensively and quantitatively evaluate the performance of all proposed methods.
Section~\ref{sec:expr:setup} elaborates the experiment setup.
Section~\ref{sec:expr:static} focuses on the accuracy evaluation on static KGs. 
We compare the evaluation efficiency and estimation quality of various methods on different static KGs with different data characteristics.
Section~\ref{sec:expr:incremental-cost} evaluates the performance of the proposed incremental evaluation methods on evolving KGs.
We simulate several typical scenarios of evolving KG evaluations in practice, and demonstrate the efficiency and effectiveness of our proposed incremental evaluation solutions.

\subsection{Experiment Setup}\label{sec:expr:setup}
\begin{table}\setlength{\tabcolsep}{2pt}
    \centering
    \small
    \caption{Data characteristics of various KGs.}
    \label{tab:dataset}
    \vspace{-1em}
    \begin{tabular}{|c|cccc|}
    \hline
     & \textbf{NELL} & \textbf{YAGO} & \textbf{MOVIE} & \textbf{\ansc{MOVIE-FULL}} \\\hline
    \makecell{Number of entities} & 817 & 822 & 288,770 & \ansc{14,495,142} \\\hline
    \makecell{Number of triples} & 1,860 & 1,386 & 2,653,870 & \ansc{
    130,591,799
    } \\\hline
    \makecell{Average cluster size\footnotemark} & 2.3 & 1.7 & 9.2 & \ansc{9.0}\\\hline
    Gold Accuracy & 91\% & 99\% & 90\% (MoE: 5\%) & \ansc{N/A} \\\hline
    \end{tabular}
    \vspace{-1em}
\end{table}

\footnotetext{Average cluster size = $\frac{\text{number of triples}}{\text{number of entities}}$.}

\subsubsection{Dataset Description}\label{sec:expr:dataset}
\ansc{We use real-life knowledge graphs, as summarized in Table~\ref{tab:dataset} and described in detail below.
}

\textbf{NELL \& YAGO}
are small sample sets drawn from the original knowledge graph \emph{NELL-Sports}~\cite{NELL,mitchell2018never} and \emph{YAGO2}~\cite{fabian2007yago,YAGO}, respectively.
\ansc{NELL is a domain-specific KG with sports-related facts mostly pertaining to athletes, coaches, teams, leagues, stadiums etc; while YAGO is not domain-specific.}
Ojha and Talukdar~\cite{ojha2017kgeval} collected manual annotated labels (true/false), evaluated by recognized workers on Amazon Mechanical Turk, for each fact in NELL and YAGO.
We use these labels as gold standard.
The ground-truth accuracies of NELL and YAGO are 91\% and 99\%, respectively.

\textbf{MOVIE},
based on IMDb\footnote{IMDb terms of service: https://www.imdb.com/conditions}~\cite{IMDB} and WiKiData~\cite{Wiki}, is a knowledge base \ansc{with entertainment-related facts mostly pertaining to actors, directors, movies, TV series, musicals etc.}
It contains more than 2 million factual triples.
To estimate the overall accuracy of MOVIE, we randomly sampled and manually evaluated 174 triples.
The unbiased estimated accuracy is 88\% within a 5\% margin of error at the 95\% confidence level.
\ansc{\textbf{MOVIE-FULL} is the full version of MOVIE, which contains more than 130 million triples. 
For cost consideration, we cannot afford manually evaluate the accuracy of MOVIE-FULL in its entirety;
we primarily use it to test scalability of the proposed methods.}

\subsubsection{Synthetic Label Generation}\label{sec:expr:syn}
Collecting human annotated labels is expensive. 
We generate a set of synthetic labels for MOVIE in order to perform in-depth comparison of different methods. 
\textbf{MOVIE-SYN} refers to a set of synthetic KGs with different label distributions.
We introduce two synthetic label generation models as follows.
\paragraph*{Random Error Model}
The probability that a triple in the KG is correct is a fixed error rate  $r_{\epsilon} \in [0,1]$.
This random error model (REM) is simple, but only has limited control over different error distributions and can not properly simulate real-life KG situations.
\vspace{-2em}
\paragraph*{Binomial Mixture Model}
Recall in Figure~\ref{fig:correlation}, we find that larger entities in the KG are more likely to have higher entity accuracy.
Based on this observation, we synthetically generate labels that better approximate such distribution of triple correctness.
First, we assume that the number of correct triples from the $i$-th entity cluster follows a \emph{binomial distribution} parameterized by the entity cluster size $M_i$ and a probability $\hat{p}_i \in [0,1]$; that is, 
$f(t) \sim \mathrm{B}(M_i, \hat{p}_i)$.
Then, to simulate real-life situations, 
we assume a relationship between $M_i$ and $\hat{p}_i$ specified by the following sigmoid-like function:
\begin{equation}\label{eq:synthetic}
    \hat{p}_i = 
    \begin{cases}
    0.5 + \epsilon, & \text{if}~M_i < k \\
    \frac{1}{1 + e^{-c(M_i - k)}} + \epsilon, & \text{if}~M_i \ge k
    \end{cases}
\end{equation}
where $\epsilon$ is a small error term from a normal distribution with mean 0 and standard deviation $\sigma$, and $c\ge 0$ scales the influence of cluster size on entity accuracy. $\epsilon$ and $c$ together control the correlation between $M_i$ and $\hat{p}_i$.

Tuning $c, \epsilon$ and $k$, the Binomial Mixture Model (BMM) allows us to experiment with different data characteristics without the burden of repeating manual annotations.
For our experiment, we use $k=3$ by default, and vary $c$ from 0.00001 to 0.5 and $\sigma$ (in $\epsilon$) from 0.1 to 1 for various experiment setting; A larger $\sigma$ and a smaller $c$ lead to a weaker correlation between the size of a cluster and its accuracy. 
By default, $c=0.01$ and $\sigma=0.1$.

\subsubsection{Cost Function}\label{sec:expr:cost}
\begin{table}
    \centering
    \caption{Manual evaluation cost (in hours) on MOVIE.}
    \label{tab:manual-evaluation}
    \begin{tabular}{|c|c|c|}
    \hline
         & SRS & TWCS ($m=10$) \\\hline
        Annotation task & \makecell{174 entities \\/ 174 triples} & \makecell{24 entities \\/ 178 triples} \\\hline
        Annotation time & 3.53 & 1.4\\\hline
        Estimation & 88\% (MoE: 4.85\%) & 90\% (MoE: 4.97\%) \\\hline
    \end{tabular}
\end{table}
\begin{figure}
    \centering
    \subfloat{\includegraphics[width=0.25\textwidth]{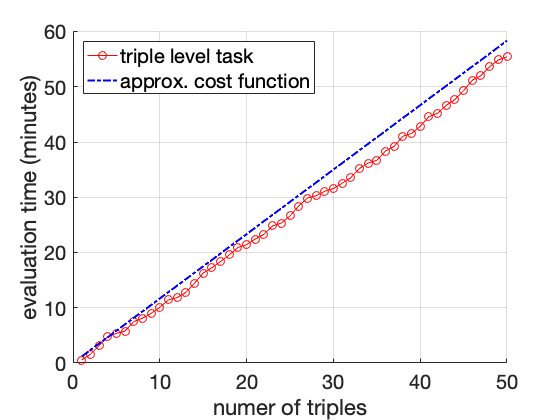}}
    \subfloat{\includegraphics[width=0.25\textwidth]{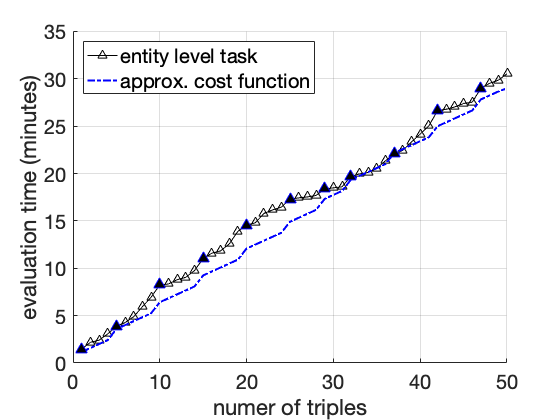}}
    \caption{Cost function fitting on different evaluation tasks.}
    \label{fig:cost-fitting}
    \vspace{-1.5em}
\end{figure}
Recall in Section~\ref{sec:cost}, we introduce an annotation model to simulate the actual manual annotation process on triple correctness, and a corresponding cost function as Eq~(\ref{eq:simplified-cost}).
We asked human annotators to do manual annotations on small portion of MOVIE selected by our proposed evaluation framework, and measured their total time usage.
Results are summarized in Table~\ref{tab:manual-evaluation}.
Then, we fit our cost function given all the data points reported in Table~\ref{tab:manual-evaluation} and Figure~\ref{fig:eval-cost}, and compute the best parameter settings as $c_1 = 45$(second) and $c_2 = 25$(second).
As shown in Figure~\ref{fig:cost-fitting}, our fitted cost function can closely approximate the actual annotation time under different types of annotation tasks.
Also, for annotation tasks in Table~\ref{tab:manual-evaluation}, the approximate cost is $174\times(45+25)/3600 \approx 3.86$ (hours) and $(24\times45 + 178\times25)/3600 \approx 1.54$ (hours), respectively, which is close to the ground-truth time usage.

Together with the synthetic label generation introduced in Section~\ref{sec:expr:syn}, 
we can effortlessly experiment and compare various sampling and evaluation methods on KGs with different data characteristics, both in terms of the evaluation quality (compared to synthetic gold standards) and efficiency (based on cost function that approximates the actual manual annotation cost).
The annotation time reported in Table~\ref{tab:manual-evaluation} and Table~\ref{tab:overall-cost} on MOVIE is the actual evaluation time usage as we measured human annotators' performance.
All other evaluation time reported in later tables or figures are approximated by the Eq~(\ref{eq:simplified-cost}) according to the samples we draw.

\subsubsection{Implementations}\label{sec:expr:baseline}

For evaluation on static KGs, we implement the Sample Collector component in our framework with \textbf{SRS} (Section~\ref{sec:srs}), \textbf{RCS} (Section~\ref{sec:rcs}), \textbf{WCS} (Section~\ref{sec:wcs}) and \textbf{TWCS} (Section~\ref{sec:2-wcs}).
TWCS is further parameterized by the second-stage sample unit size $m$.
If $m$ is not specified, we run TWCS with the optimal choice of $m$. 
We also implement TWCS with two stratification strategies (Section~\ref{sec:stratified}).
\textbf{Size Stratification} method first stratifies the clusters by their size using the \emph{Cumulative Square root of Frequency} (Cumulative $\sqrt{F}$)~\cite{dalenius1959minimum}, and then applies the iterative stratified TWCS as introduced in Section~\ref{sec:stratified}.
Since we have the ground-truth labels for YAGO, NELL and MOVIE-SYN, \textbf{Oracle Stratification} method directly stratifies the clusters by their entity accuracy, which is the perfect stratification but not possible in practice though.
The evaluation time reported by oracle stratification can be seen as the lower bound of cost, showing the potential improvement of TWCS with stratification.
\ansc{We consider \textbf{KGEval}, proposed in \cite{ojha2017kgeval}, as another baseline solution, and compare its performance with our best solution on the same KGs: YAGO and NELL.}

For evaluation on evolving KGs, since there are
no well established baseline solutions, we consider a simple \textbf{Baseline} of independently applying TWCS on each snapshot of evolving KGs.
Besides that, we implement \textbf{RS}, incremental evaluation based on Reservoir Sampling (Section~\ref{sec:reservoir-incremental}), and \ansb{\textbf{SS}, incremental evaluation based on stratified sampling (Section~\ref{sec:stratified-incremental}). }

All methods were implemented in Python3,
and all experiments were performed on a Linux machine with two
Intel Xeon E5-2640 v4 2.4GHz processor with 256GB of memory.
\subsubsection{Performance Evaluation Metric}\label{sec:expr:metric}
Performance of various methods are evaluated using the following two metrics: sample size (number of triples/entities in the sample) and manual annotation time.
By default, all evaluation tasks (on static KGs and evolving KGs) are configured as MoE less than 5\% with 95\% confidence level ($\alpha=5\%$).
Considering the uncertainty of sampling approaches, we repeat each method 1000 times and report the average sample size/evaluation time usage with standard deviation.
\ansmeta{To demonstrate the unbiasedness of our evaluation results, we also report the average accuracy estimation with standard deviation under 1000 random trials.}

\begin{table*}
\begin{minipage}{0.6\textwidth}
    \centering
    \setlength{\tabcolsep}{1.5pt}
    \small
    \begin{threeparttable}
    \caption{
    Performance comparison of various solutions on static KGs.\label{tab:overall-cost}
    }
    \vspace{-1em}
    \begin{tabular}{|c|c|c|c|c|c|c|}
    \cline{2-7}
    \multicolumn{1}{c|}{} & \multicolumn{2}{c|}{\makecell{MOVIE\\(gold acc. 90\%, 5\% MoE)}} & \multicolumn{2}{c|}{\makecell{NELL\\(gold acc. 91\%)}} & \multicolumn{2}{c|}{\makecell{YAGO\\(gold acc. 99\%)}} \\\cline{2-7}
    \multicolumn{1}{c|}{} & \makecell{Annotation \\Time(hours)
    } & \ansmeta{Estimation} &  \makecell{Annotation\\Time (hours)} & \ansmeta{Estimation} &  \makecell{Annotation \\Time(hours)} & \ansmeta{Estimation}\tnote{\ddag} \\\hline
    SRS & 3.53\tnote{*} & \ansmeta{$90\%$} & 2.3$\pm$0.45 & \ansmeta{$91.5\%\pm 2.1\%$} &0.45$\pm$0.17 & 
    \makecell{\ansmeta{99.6\%} \\ \ansmeta{(96.7\%-100\%)}}
    \\\hline
    RCS & $>5$\tnote{*} 
    & \ansmeta{95\%}\tnote{\dag}
    & 8.25$\pm$2.55 & \ansmeta{$90.5\%\pm 2.4\%$} & 10$\pm$0.56 & 
    \makecell{\ansmeta{98.9\%} \\ \ansmeta{(95.3\%-100\%)}}
    \\\hline
    WCS & $>5$\tnote{*} 
    & \ansmeta{93\%}\tnote{\dag} 
    & 1.92$\pm$0.62 & \ansmeta{$91.6\%\pm 2.3\%$} & 0.49$\pm$0.04 & 
    \makecell{\ansmeta{99.2\%} \\ \ansmeta{(96.7\%-100\%)}}
    \\\hline
    TWCS & \cellcolor{blue!25}1.4\tnote{*} & \ansmeta{$88\%$} & \cellcolor{blue!25}1.85$\pm$0.6 & \ansmeta{$91.6\%\pm 2.2\%$} & \cellcolor{blue!25}0.44$\pm$0.07 & 
    \makecell{\ansmeta{99.2\%} \\ \ansmeta{(96.7\%-100\%)}}
    \\\hline
    \end{tabular}
    \begin{tablenotes}
     \item[*] Actual manual evaluation cost; other costs are estimated using Eq(\ref{eq:simplified-cost}) and averaged over 1000 random runs.
     \item[\dag] For economic considerations, we stop manual annotation process at 5 hours for RCS and WCS. 
     Note that estimations in these two cases (95\% and 93\%) do not satisfy the 5\% MoE with 95\% confidence level.
    \end{tablenotes}
    \end{threeparttable}
    \end{minipage}
\begin{minipage}{0.38\textwidth}
\setlength{\tabcolsep}{1.5pt}
 \centering
    \small
    \begin{threeparttable}
    \caption{\ansmeta{Performance comparison of TWCS and KGEval on NELL and YAGO.}\label{tab:kgeval}}
     \vspace{-1em}
    \begin{tabular}{|c|c|c|c|c|}
    \cline{2-5} 
    \multicolumn{1}{c|}{} & \multicolumn{2}{c|}{\makecell{NELL\\(gold acc. 91\%)}} & \multicolumn{2}{c|}{\makecell{YAGO\\(gold acc. 99\%)}} \\\cline{2-5}
    \multicolumn{1}{c|}{} & \multirow{2}{*}{KGEval} & \multirow{2}{*}{TWCS} & \multirow{2}{*}{KGEval} & \multirow{2}{*}{TWCS} \\\hhline{~~~~~}
    \multicolumn{1}{c|}{} &  &  &  &  \\\hline
    \makecell{Machine Time\\(sample generation)} & \makecell{12.44\\hours} & \cellcolor{blue!25}$<$1 second & \makecell{18.13\\hours} & \cellcolor{blue!25}$<$1 second \\\hline
    \makecell{\# of triples\\annotated} & 140 & $149\pm 47$ & 204 & 32$\pm$5 \\\hline
    \makecell{Annotation\\Time (hours)} & 2.3 & \cellcolor{blue!25}1.85$\pm$0.6 & 3.17 & \cellcolor{blue!25}0.44$\pm$0.07 \\\hline
    \makecell{\ansmeta{Estimation}} & \ansmeta{91.84\%} & \makecell{\ansmeta{91.63\%}\\ \ansmeta{$\pm$2.3\%}} & \ansmeta{99.30\%} & 
    \makecell{\ansmeta{99.2\%}\tnote{\ddag} \\ \ansmeta{(96.7\%-100\%)}}
    \\\hline
    \end{tabular}
    \begin{tablenotes}
     \item[\ddag] For the highly accurate KG YAGO, we report \emph{empirical confidence interval} instead of mean and standard deviation. Since accuracy is always capped at 100\%, empirical confidence interval can better represent the data distribution in this case.
    \end{tablenotes}
    \end{threeparttable}
\end{minipage}
\end{table*}
    
\begin{figure}
\centering
\subfloat[Sample size as confidence level $1-\alpha$ varies.]{\includegraphics[width=0.48\textwidth]{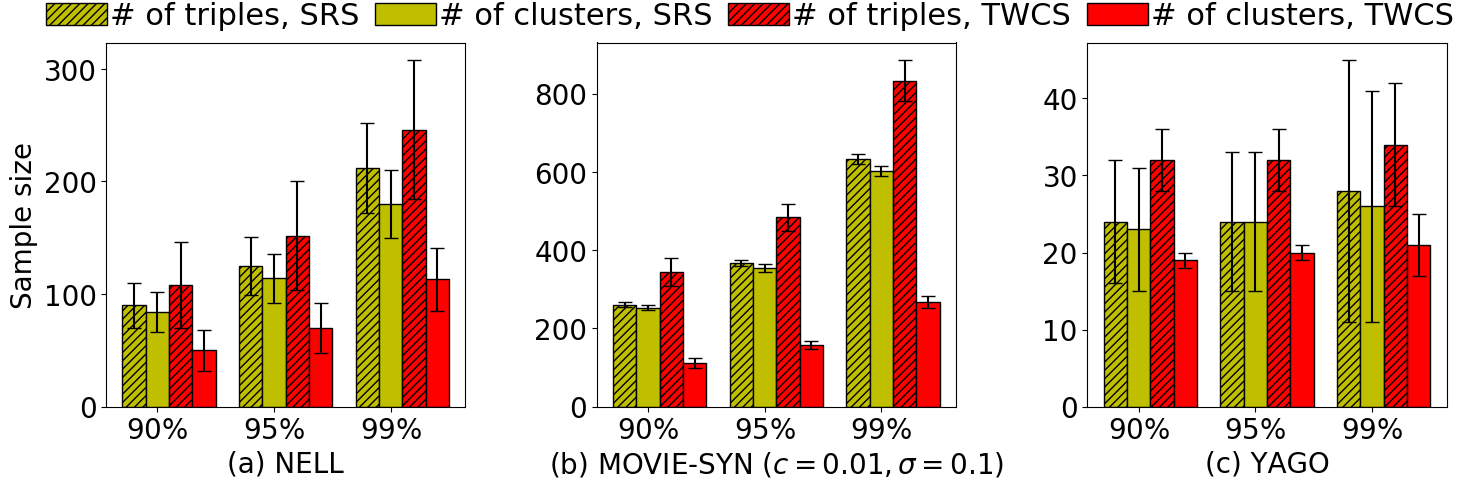}}\\
\subfloat[Evaluation time as confidence level $1-\alpha$ varies (reduction ratio shown on top of bars).]{\includegraphics[width=0.48\textwidth]{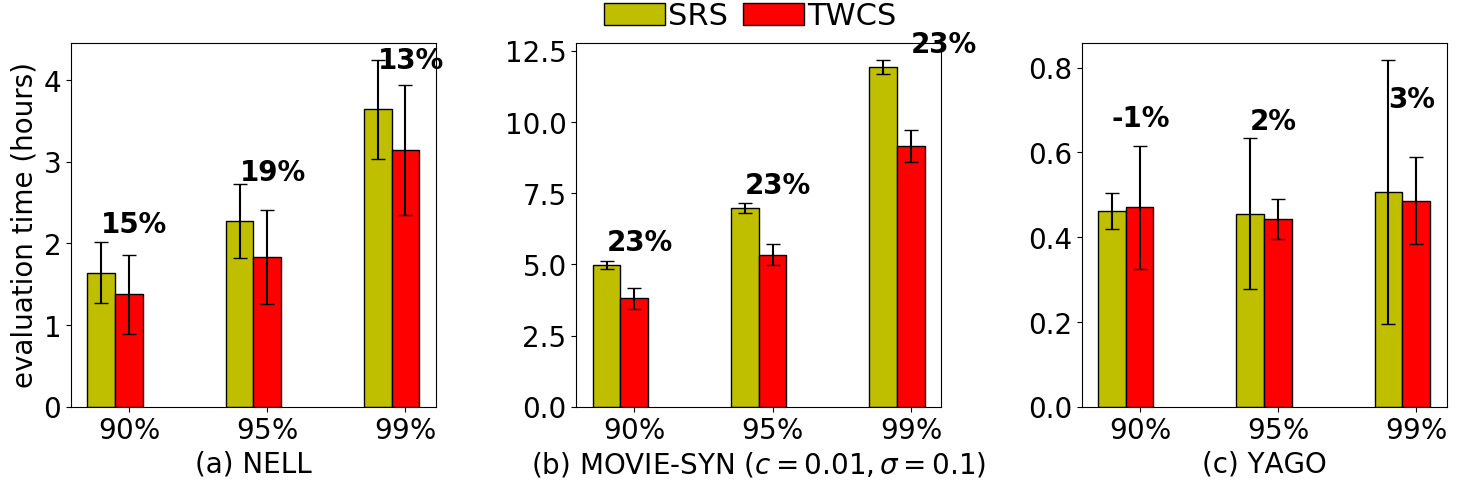}}
\vspace{-1em}
\caption{Performance comparison of SRS and TWCS on static KGs.\label{fig:cost-vary-alpha}}
\vspace{-1.5em}
\end{figure}

\subsection{Evaluation on Static KG}\label{sec:expr:static}
In this section, we compare five methods for evaluation on static KGs: SRS, RCS, WCS, TWCS and KGEval~\cite{ojha2017kgeval}.
\ansc{
Since by default we fix the level of precision (i.e., desired confidence interval) of estimation result, we focus on evaluation efficiency.
}
\subsubsection{Evaluation Efficiency}
\paragraph*{TWCS vs. All}\label{sec:expr:WCS-efficiency}
We start with an overview of performance comparison of various solutions on evaluation tasks on Static KGs.
Results are summarized in Table~\ref{tab:overall-cost}.
Best results in evaluation efficiency are colored in blue. 
We highlight the following observations.
First, TWCS achieves the lowest evaluation cost across different KGs, speeding the evaluation process up to 60\% (on MOVIE with actual human evaluation cost), without evaluation quality loss.
TWCS combines the benefits of weighted cluster sampling and multi-stage sampling. It lowers entity identification costs by sampling triples in entity groups and applies the second-stage sampling to cap the per-cluster annotation cost.
As expected, TWCS is the best choice for efficient KG accuracy evaluation overall.
\ansmeta{
Furthermore, as shown in the Estimation column of Table~\ref{tab:overall-cost}, all four proposed solutions provide unbiased accuracy estimations with small ($< 3\%$) deviation from ground-truth accuracy.
}

\ansmeta{
\paragraph*{TWCS vs.\ KGEval} We further compare the performance of our best solution TWCS with KGEval on NELL and YAGO.
See Section~\ref{sec:related-work} for a detailed description of KGEval.
As shown in Table~\ref{tab:kgeval}, TWCS could significantly improve the evaluation process both in machine time (up to 10,000$\times$ speedup) and manual annotation cost (up to 80\% cost reduction) without estimation quality loss.
Due to KGEval's scalability issue, we also find it infeasible to apply KGEval on large-scale KGs.
In addition, TWCS  gives unbiased estimation with user-required confidence interval, while KGEval does not have such a feature.

}

\paragraph*{TWCS vs.\ SRS} Since SRS is the only solution that has comparable performance to TWCS, we dive deeper into the comparison of SRS and TWCS.
Figure~\ref{fig:cost-vary-alpha} shows the comparison in terms of sample size and annotation cost on all three KGs with various evaluation tasks.
We summarize the important observations as follows.
First, in Figure~\ref{fig:cost-vary-alpha}-1, TWCS draws fewer entity clusters than SRS does, even though the number of triples annotated in total by TWCS is slightly higher than that of SRS.
Considering that the dominant factor in annotation process is of entity identification, TWCS still saves a fair amount of annotation time.
Second, Figure~\ref{fig:cost-vary-alpha}-2 quantifies the cost reduction ratio (shown on top of the bars) provided by TWCS.
Simulation results suggest that TWCS outperforms SRS by a margin up to 20\% on various evaluation tasks with different confidence level on estimation quality and across different KGs.
Even on the highly accurate YAGO, TWCS still can save 3\% of the time compare to SRS when the estimation confidence level is 99\%.
\ansa{
It is worth mentioning that on highly accurate KGs, like YAGO with 99\% accuracy, there is no notable performance difference between TWCS and SRS.
Figure~\ref{fig:cost-vary-alpha}-1-c shows that both methods require only 20 to 30 triples to get accurate estimations.
In such case, sampling individual triples or sampling triples grouped by their subjects does not differ much in terms of manual annotation cost.
In fact, when the evaluation task only requires a few triples, the annotation overhead of TWCS makes it potentially less efficient than SRS, which explains why TWCS gives a negative reduction ratio at $90\%$ confidence level for YAGO in Figure~\ref{fig:cost-vary-alpha}-2-c.
}
\vspace{-1em}
\begin{figure*}
\centering
\includegraphics[width=0.9\textwidth]{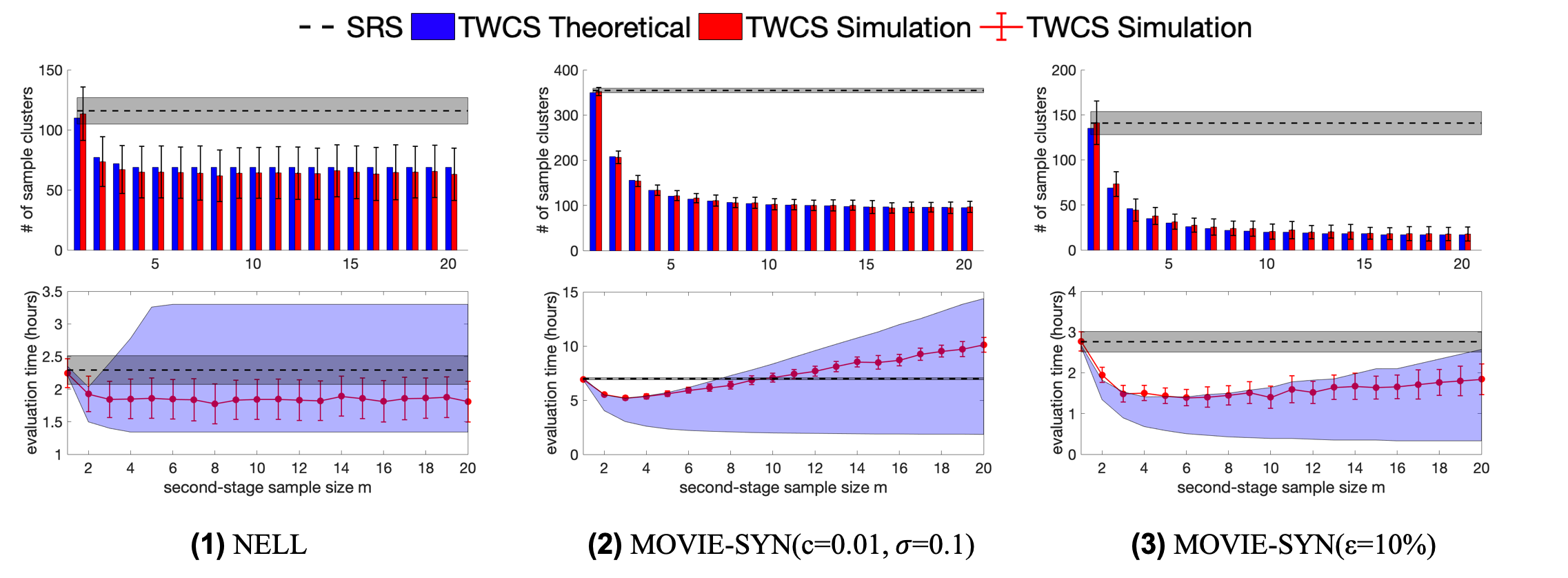}
\caption{
\ansc{Finding optimal second stage sample size $m$ on NELL and MOVIE-SYN with various synthetic generated labels. Grey ribbon and Blue ribbon in plots represent the standard deviation of SRS performances and the theoretical range of TWCS annotation cost respectively.}
\label{fig:optimal-m-varing}}
\vspace{-1.5em}
\end{figure*}

\subsubsection{Optimal Sampling Unit Size of TWCS}\label{sec:expr:optimal-m}
So far, we run experiments using TWCS with the best choice of second-stage sample size $m$. 
In this section, we discuss the optimal value of $m$ for TWCS and provide some guidelines on how to choose the (near-) optimal $m$ in practice.
We present the performance of TWCS on NELL and two instances of MOVIE-SYN as the second-stage sampling size $m$ varies from 1 to 20 in Figure~\ref{fig:optimal-m-varing}, using SRS as a reference.
\ansc{Since the ground-truth labels are available, we also compare the theoretical results (blue ribbon with upper/lower bound in Figure~\ref{fig:optimal-m-varing}) based on Eq~(\ref{eq:wcs-var-final}) with the simulation results.}
For each setting, numbers are reported averaging over 1K random runs. \ansc{Standard deviation is shown as error bar and grey ribbon in all plots in Figure~\ref{fig:optimal-m-varing}.}

For sample size comparison:
First, when $m=1$, TWCS is equivalent to SRS (recall Proposition~\ref{lemma:wcs-to-srs}), so the sample size (and evaluation time) reported by TWCS is very close to SRS.
Second, as $m$ increases, the sample cluster size would first drop significantly and then quickly hit the plateau, showing that a large value of $m$ does not help to further decrease the number of sample clusters.

\ansc{For annotation time, the theoretical results are shown as blue ribbon with the upper bound (assuming all selected sample clusters are larger than $m$) and the lower bound (assuming all selected sample clusters have size of 1).}
Again, when $m=1$, the evaluation cost of SRS and TWCS are roughly the same.
Then, on two instances of MOVIE-SYN (Figure~\ref{fig:optimal-m-varing}-2 and Figure~\ref{fig:optimal-m-varing}-3), the annotation time decreases as $m$ increases from 1 to around 5 and then starts to go up, which could be even higher than SRS when $m\geq10$ (see Figure~\ref{fig:optimal-m-varing}-2).
A larger value of $m$ potentially leads to more triples to be annotated, as it can not further reduce the number of sample clusters but we are expected to choose more triples from each selected cluster.
On NELL, things are little different: the annotation time drops as $m$ increases from 1 to around 5 but then roughly stays the same.
That is because NELL has a very skewed long-tail distribution on cluster size - more than 98\% of the clusters have size smaller than 5.
Hence, when $m$ becomes larger than 5 and no matter how large the $m$ will be, the total number of triples we evaluated is roughly a constant and it is not affected by $m$ anymore.

In sum, there is an obvious trade-off between $m$ and evaluation cost, as observed in all plots Figure~\ref{fig:optimal-m-varing}.
The optimal choices on $m$ across KGs (with different data characteristics and  accuracy distributions) seems all fall into a narrow range between 3 to 5.
However, entity accuracy distributions influence the relative performance of TWCS compared to SRS.
For instance, on MOVIE-SYN with $\epsilon=10\%$, the synthetic label generations make the cluster accuracy among entities more similar (less variance among accuracy of entity clusters), so we can see TWCS beats SRS by a wide margin up to 50\%. 
In contrast, for other label distributions, the cost reductions provided by TWCS are less than 50\%.
Considering all results shown in Figure~\ref{fig:optimal-m-varing}, TWCS, by carefully choosing the second-stage sample size $m$, always outperforms SRS.

To conclude this section, the optimal choice of $m$ and the performance of TWCS depend both on cluster size distribution and cluster accuracy distribution of the KG.
In practice, we do not have all such information beforehand.
However, as a practical guideline, we suggest that $m$ should not be too large.
We find a small range of $m$ from roughly 3-5 to give the (near-) minimum evaluation cost of TWCS for all KGs considered in our experiments.

\subsubsection{TWCS with Stratification}\label{sec:expr:stratification}
\begin{table*}
    \centering
    \begin{threeparttable}
    \caption{Evaluation cost (hours) by TWCS with stratification using cumulative $\sqrt{F}$; NELL has two strata and MOVIE/MOVIE-SYN has four strata.
    A good stratification strategy could further boost the efficienty of basic TWCS up to 40\%.}
    \label{tab:stratified-sampling}
    \begin{tabular}{|c|c|c|c|c|c|c|}
        \cline{2-7}
    \multicolumn{1}{c|}{} & \multicolumn{2}{c|}{\makecell{NELL\\(gold acc. 91\%)}} & \multicolumn{2}{c|}{\makecell{MOVIE-SYN\\($c=0.01,\sigma=0.1$, gold acc. 62\%)}} & \multicolumn{2}{c|}{\makecell{MOVIE\\(gold acc. 90\%)}} \\\cline{2-7}
    \multicolumn{1}{c|}{} & \makecell{Annotation Cost \\(hours)} & \ansmeta{Estimation} & \makecell{Annotation Cost \\(hours)} & \ansmeta{Estimation} & \makecell{Annotation Cost \\(hours)} & \ansmeta{Estimation} \\\hline
    SRS & 2.3$\pm$0.45 & \ansmeta{91.5\%$\pm$2.1\%} & 6.99$\pm$0.1 & \ansmeta{61.7\%$\pm$2\%} & 3.53\tnote{*} & \ansmeta{90\%} \\\hline
    TWCS & 1.85$\pm$0.6 & \ansmeta{91.6\%$\pm$2.2\%} & 5.25$\pm$0.46 & \ansmeta{62\%$\pm$2.3\%} & 1.4\tnote{*} & \ansmeta{88\%} \\\hline
    TWCS w/ Size Stratification & 1.90$\pm$0.53 & \ansmeta{91.9\%$\pm$2.3\%} & 3.97$\pm$0.5 & \ansmeta{61.8\%$\pm$2\%} & 1.3\tnote{*} & \ansmeta{88\%} \\\hline
    TWCS w/ Oracle Stratification & 1.04$\pm$0.06 & \ansmeta{91.4\%$\pm$2.4\%} & 2.87$\pm$0.3 & \ansmeta{61.5\%$\pm$2\%} & \ansc{N/A}\tnote{\dag} & \ansc{N/A}\tnote{\dag} \\\hline
    \end{tabular}
    \begin{tablenotes}
     \item[*] Actual manual evaluation cost; other evaluation costs are estimated using Eq(\ref{eq:simplified-cost}) and averaged over 1000 random runs.
     \item[\dag] Since we do not collect manually evaluated labels for all triples in MOVIE, oracle stratification is not applicable here.
    \end{tablenotes}
    \end{threeparttable}
    \vspace{-1.5em}
\end{table*}
Previous sections already demonstrate the efficiency of TWCS.
Now, we further show that our evaluation framework could achieve even lower cost with proper stratification over entity clusters in KG.
Recall that size stratification partitions clusters by the size while oracle stratification stratifies clusters using the entity accuracy.
Table~\ref{tab:stratified-sampling} lists and compares the evaluation cost of different methods on NELL, MOVIE-SYN ( $c=0.01,\sigma=0.1$) and MOVIE.

On MOVIE-SYN, we observe that 
TWCS with stratification over entity clusters
could further significantly reduce the annotation time. Compared to SRS, size stratification speeds up the evaluation process up to 40\% (20\% reduction compared to TWCS without stratification) and oracle stratification can make the annotation time less than 3 hours.
Recall that MOVIE-SYN has synthetic labels generated by BMM introduced in Section~\ref{sec:expr:syn}.
We explicitly correlate entity accuracy with entity cluster size using Eq~(\ref{eq:synthetic}).
Hence, a simple strategy as size stratification could already do a great job to group clusters with similar entity accuracies, reduce overall variance and boost the efficiency of basic TWCS.

However, on NELL and MOVIE, we observe that applying stratification on TWCS does not help too much and it might be even slightly worse than the basic TWCS (as on NELL).
This is because, in practice, cluster size may serve as a good signal indicating similarities among entity accuracies for large clusters but not for those small ones, and the overall variance is not reduced as we expected.
However, the oracle stratification tells us that the cost can potentially be as low as about an hour on NELL.
If we use good stratification strategies in term of grouping clusters with similar entity accuracy together, we are expected to achieve lower cost that close to the oracle evaluation cost.
\ansmeta{Again, TWCS with stratification still provides unbiased estimation of overall accuracy, as confirmed by numbers reported in Estimation column of Table~\ref{tab:stratified-sampling}}.
\begin{figure}
    \centering
    \subfloat[Varying KG size.]{\includegraphics[width=0.25\textwidth]{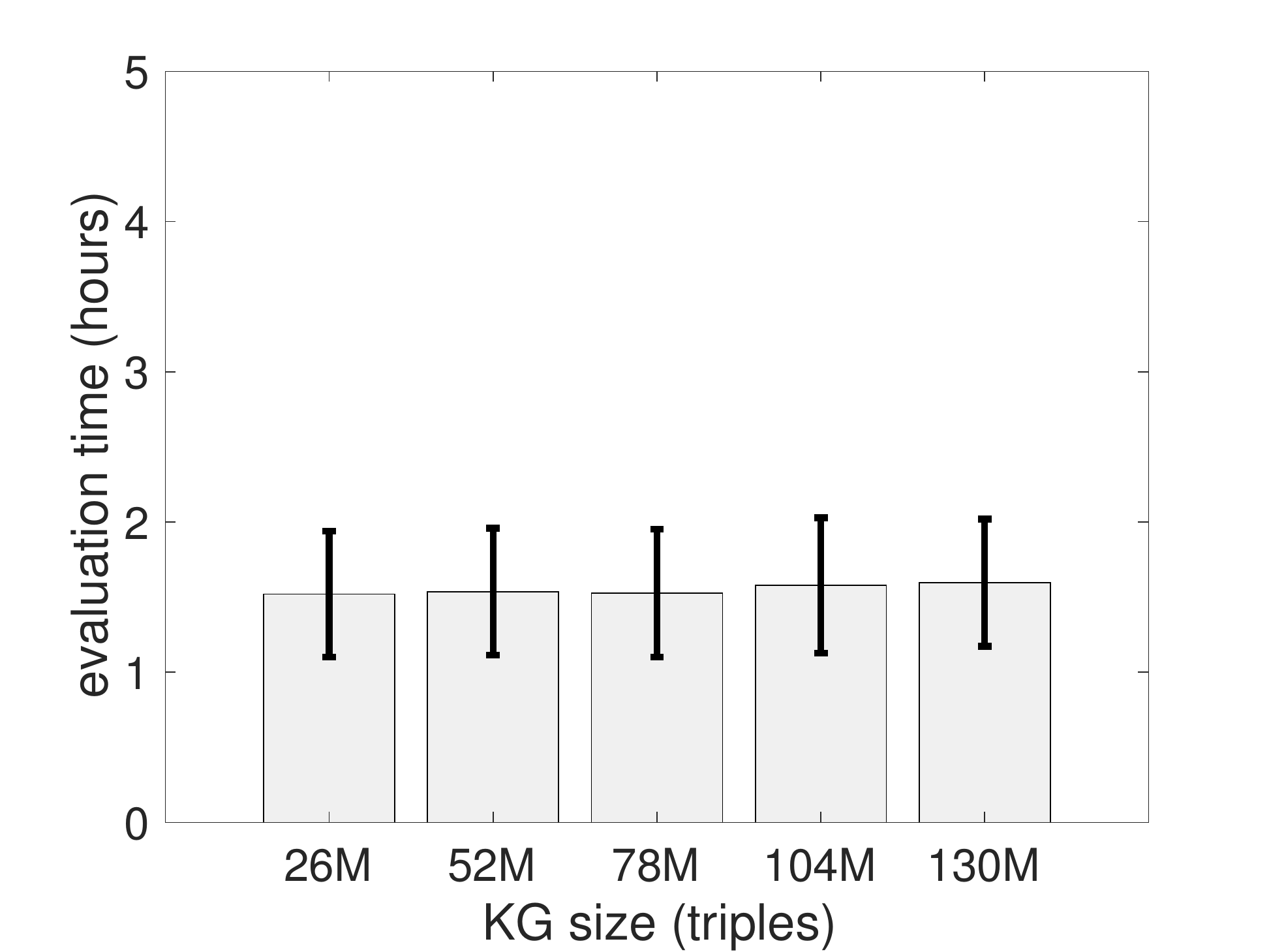}}
    \subfloat[Varying accuracy.]{\includegraphics[width=0.25\textwidth]{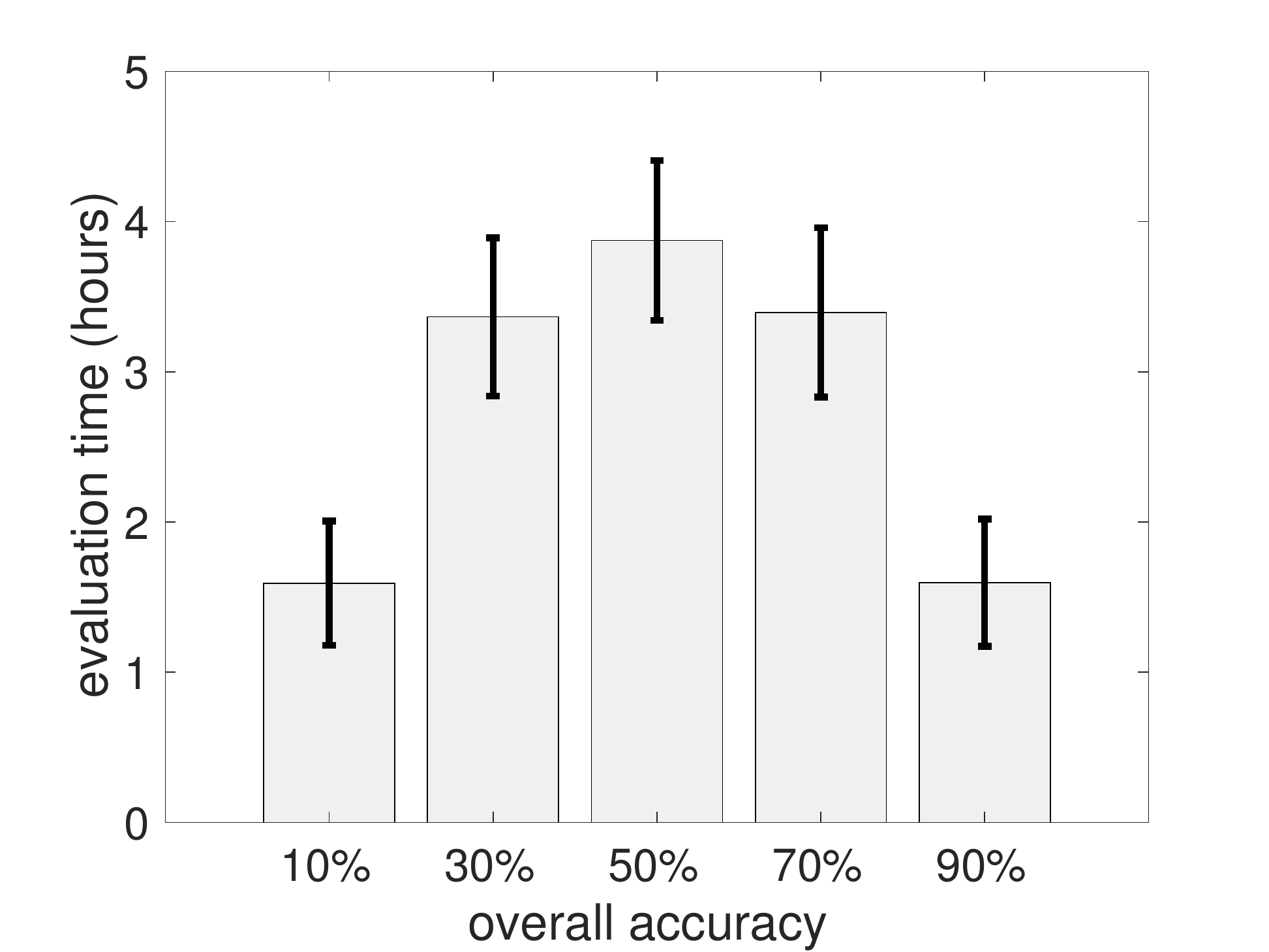}}
    \caption{\ansc{TWCS evaluation cost on various KGs with different size and different overall accuracy.}}
    \label{fig:scalability}
    \vspace{-1.5em}
\end{figure}

\subsubsection{Scalability of TWCS}\label{sec:expr:scalability}

\ansc{
Our best solution, TWCS, is based on sampling theory. 
The most appealing property of sampling is that the sample size required to achieve a certain level of precision is largely unaffected by the underlying population size, provided that the underlying population  is  large  enough.
Hence, TWCS is highly scalable over large KGs.
To test the scalability of TWCS, we choose a set of sample KGs drawn from MOVIE-FULL with the number of triples ranging from 26 million to 130 million (full size of MOVIE-FULL). 
For simplicity, we synthetically generate labels for these sample KGs using REM with $r_\epsilon = 0.1$, which in turn fixes the overall accuracy of all KGs to 90\%.
All results reported are averaged over 1K random runs with error bars showing the standard deviation.
As shown in Figure~\ref{fig:scalability}-1, the evaluation cost of TWCS roughly stays nearly the same as KG size grows, even on the full size of MOVIE-FULL with 130 million triples.
On the other hand, we vary the overall accuracy from 10\% to 90\% on MOVIE-FULL and report the evaluation cost in Figure~\ref{fig:scalability}-2.
The cost peaks at 50\% accuracy, where the variance among triples' correctness (1 or 0) reaches the maximum.

To summarize, the TWCS evaluation cost is affected by the underlying KG accuracy (and reaches the highest when accuracy is around 50\%), but this cost is not sensitive to the KG size. 

}

\subsection{Incremental Evaluation on Evolving KG}\label{sec:expr:incremental-cost}
\begin{figure}
    \centering
    \subfloat[Varying update size.]{\includegraphics[width=0.25\textwidth]{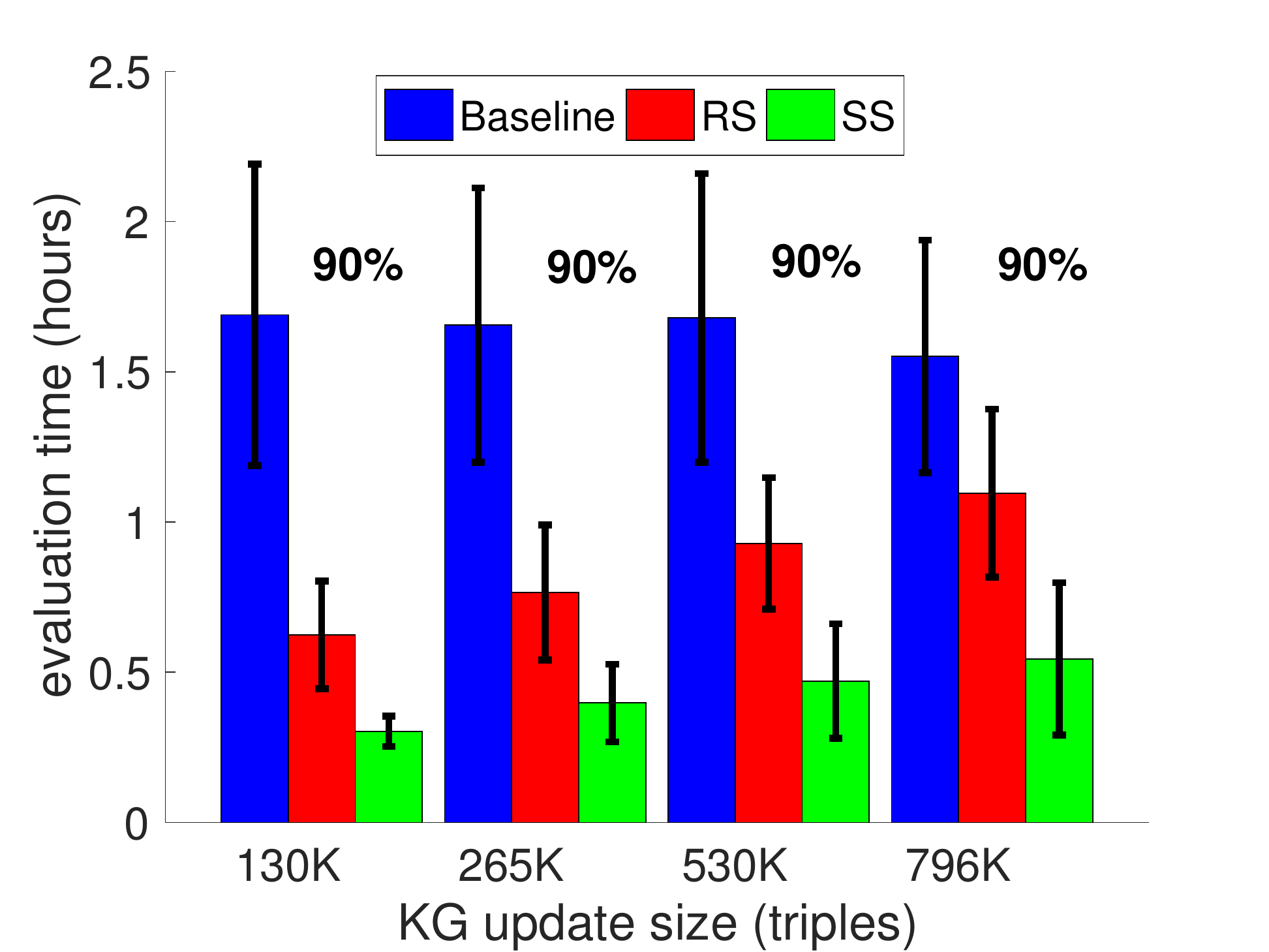}}
    \subfloat[Varying accuracy.]{\includegraphics[width=0.25\textwidth]{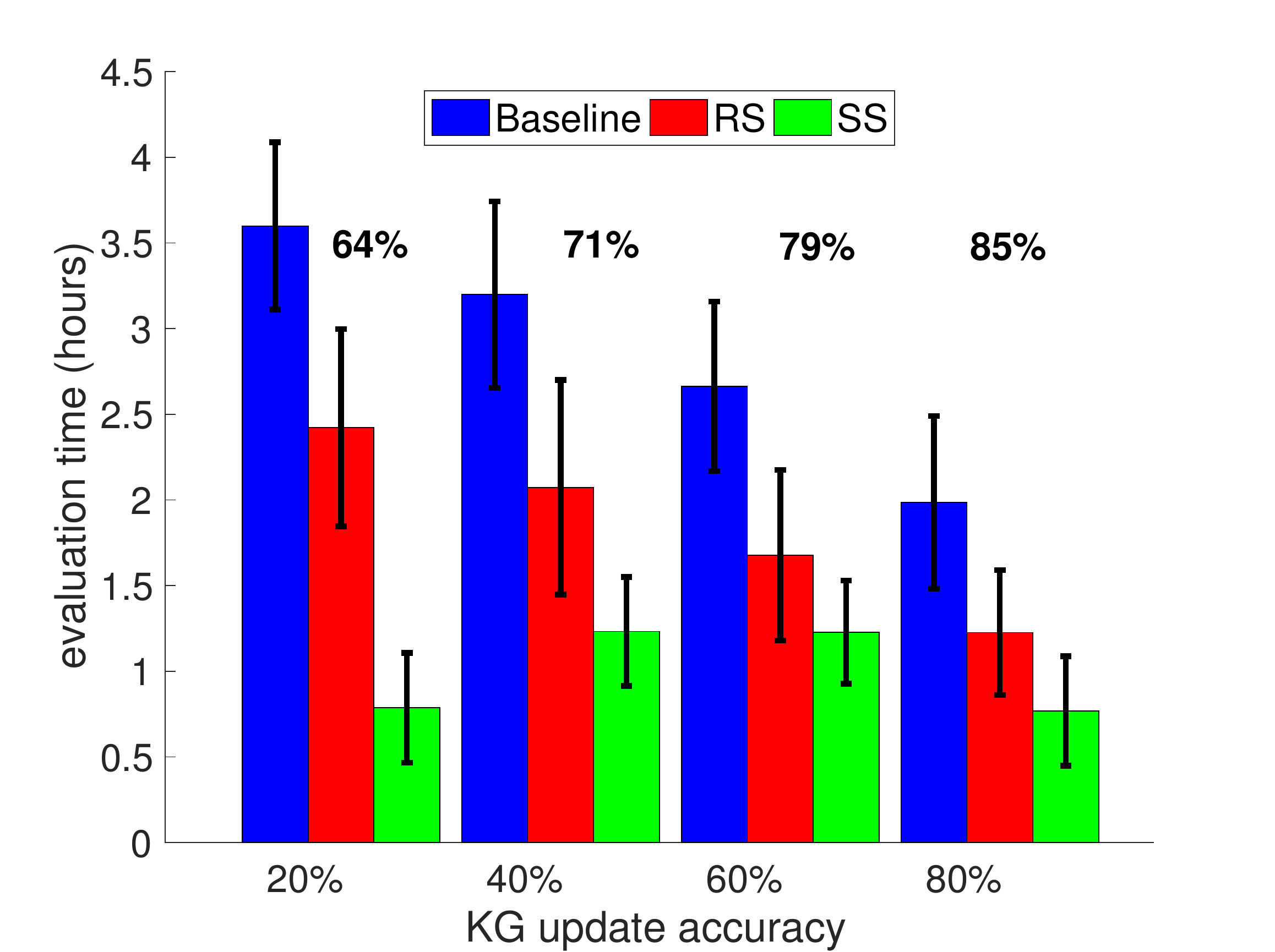}}
    \vspace{-0.5em}
    \caption{\ansmeta{Comparing evaluation cost for various solutions on evolving KG with a single update batch. Overall accuracy after update is shown on top of bars.}}
    \label{fig:kg-single-update-cost}
    \vspace{-1.5em}
\end{figure}
In this section, we present an in-depth investigation on evolving KG evaluations.
First, we set the base KG to be a 50\% subset randomly selected from MOVIE.
Then, we draw multiple batches of random sets from MOVIE-FULL as KG updates.
This setting better approximates the evolving KG behavior in real-life applications, as KG updates consist of both introducing new entities and enriching existing entities.
Our evaluation framework can handle both cases.
Since the gold accuracy of MOVIE is about 90\%, we synthetically generate labels for the base KG using REM with $r_\epsilon = 0.1$, which also gives an overall accuracy around 90\%.

\subsubsection{Single Batch of Update}\label{sec:expr:single-update}
We start with a single update batch to the base KG to understand comparison of the proposed solutions.

In the first experiment, we fix the update accuracy at 90\%, and vary the update size (number of triples) from 130K ($\sim10\%$ of base KG) to 796K ($\sim50\%$ of base KG).
Figure~\ref{fig:kg-single-update-cost}-1 shows the comparison of annotation time of three solutions.
The Baseline performs the worst because it discards the annotation results collected from previous round of evaluation and applies static evaluation from scratch.
For RS, recall from Proposition~\ref{lemma:incremental-cost-bound} that the expected number of new triples replacing annotated triples in the reservoir would increase as the size of update grows; hence, the corresponding evaluation cost also goes up as applying larger KG update.%
\ansb{
SS, based on stratified sampling, keeps all annotated triples from the previous rounds of evaluation, thus gives the lowest evaluation cost.
The cost of SS also slowly increases as KG update size increases, 
because a larger KG update makes its corresponding stratum constitute a larger weight among all strata, requiring more samples in this stratum to further reduce its own variance.
We can see in Figure~\ref{fig:kg-single-update-cost}-1 that SS further reduces the annotation cost by about 50\% compared to RS.
}

In the second experiment, we fix the update size at 796K triples, and vary the update accuracy from 20\% to 80\%.
Note in this case, after applying the update, the overall accuracy also changes accordingly.
Evaluation costs of all three methods are shown in Figure~\ref{fig:kg-single-update-cost}-2.
It is not surprising to see that Baseline performs better as KG update is more accurate (or more precisely, the overall KG accuracy after applying update is more accurate).
RS also performs better when update is more accurate. 
Even though we fix the update size, which makes the number of new triples inserted into the reservoir roughly remains the same, as overall KG is more accurate, we still can expect to annotate less additional triples to reduce the variance of estimation after sample update.
\ansb{
Lastly, the evaluation cost of SS depends on the accuracy of the KG update: it is more expensive when the update accuracy is close to 50\%, and less when update accuracy is close to 0\% and 100\%.
This observation also echoes Figure~\ref{fig:scalability}-2, showing that a highly accurate KG requires fewer samples to produce an estimation with low variance.
Overall, SS still outperforms RS with cost reduction ratios ranging from 20\% to 67\%.
}

To conclude this section, incremental evaluation methods, RS and SS, are more efficient on evolving KGs.
RS depends both on update size and overall accuracy, while \ansb{SS is relatively independent on update size and more impacted by update accuracy.
In terms of efficiency of evolving KG evaluation, SS is the clear winner.}

\begin{figure}
    \centering
    \subfloat[\ansb{Estimation average over 1K runs.}]{\includegraphics[width=0.4\textwidth]{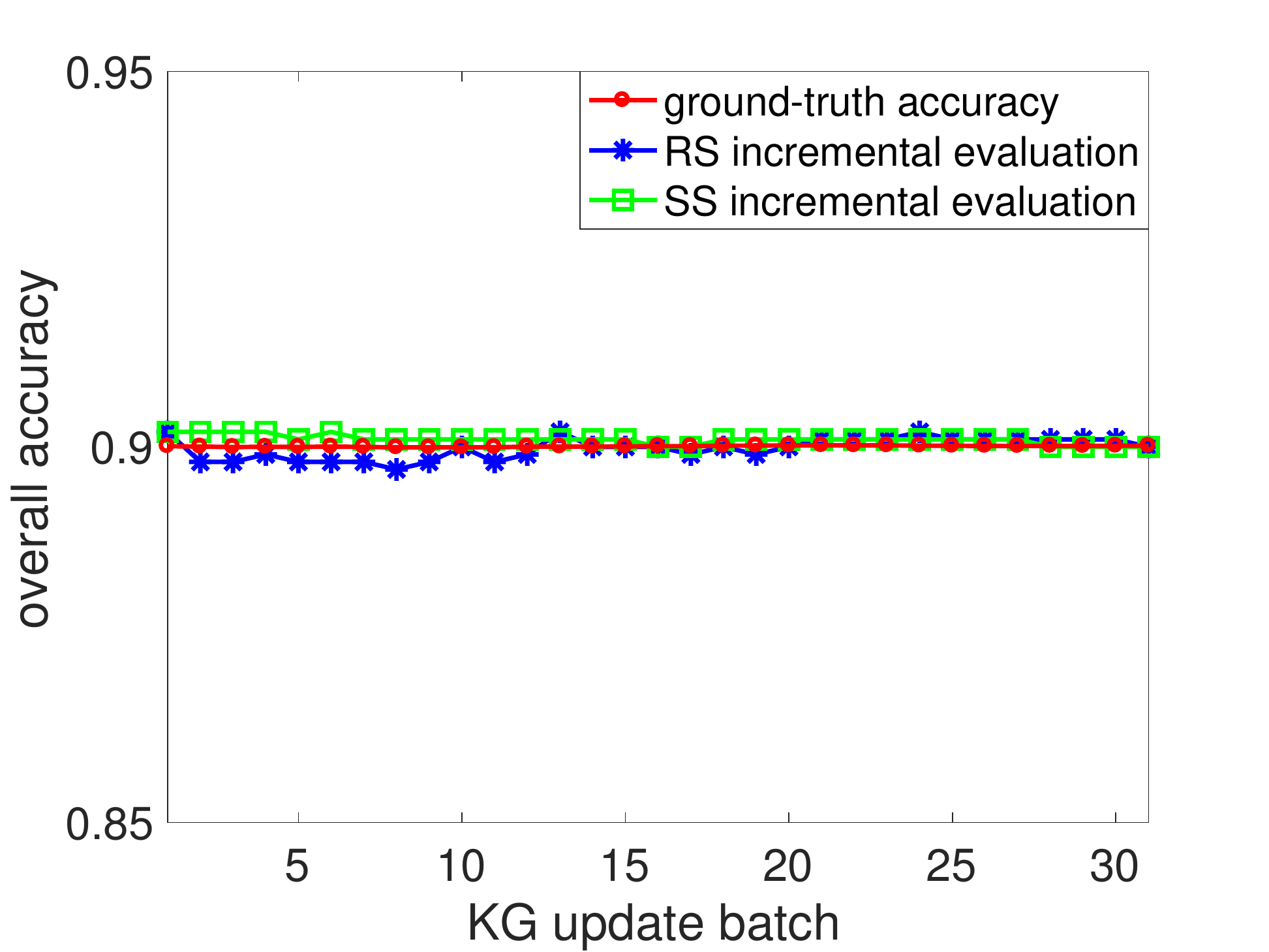}}\\
    \subfloat[\ansb{One run starts with over-estimation.}]{\includegraphics[width=0.235\textwidth]{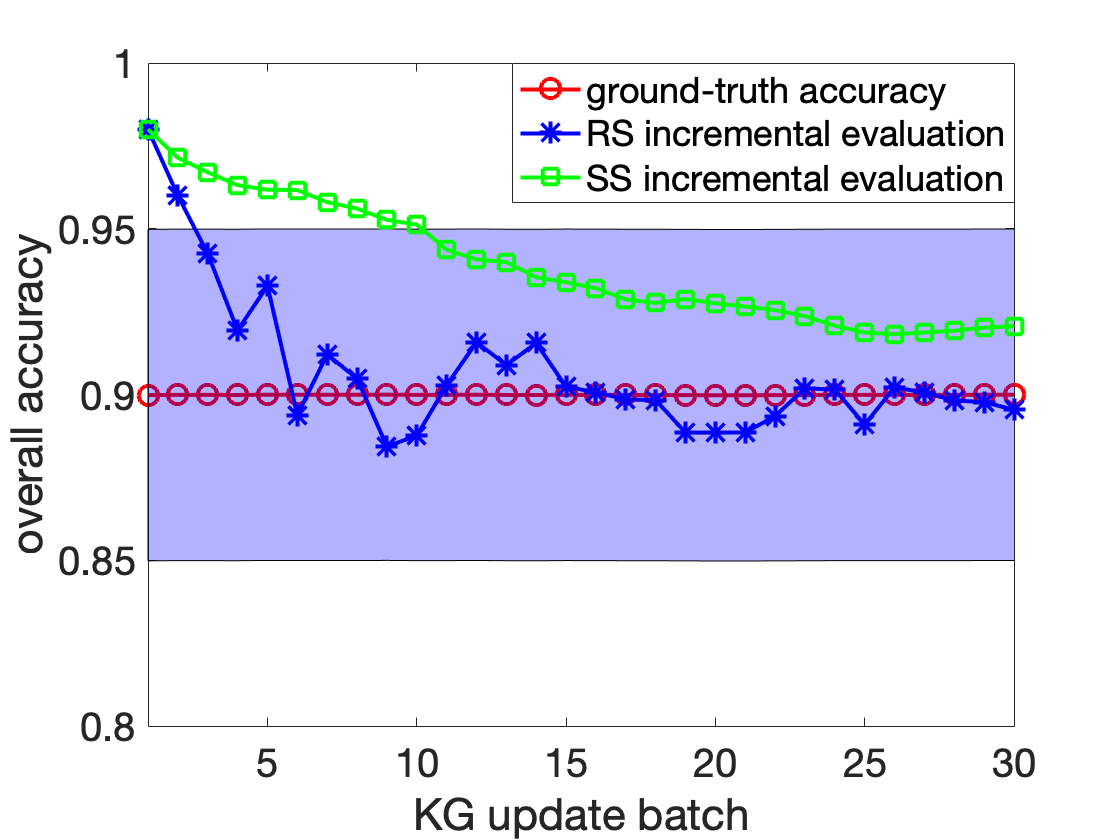}}\hfill
    \subfloat[\ansb{One run starts with under-estimation.}]{\includegraphics[width=0.235\textwidth]{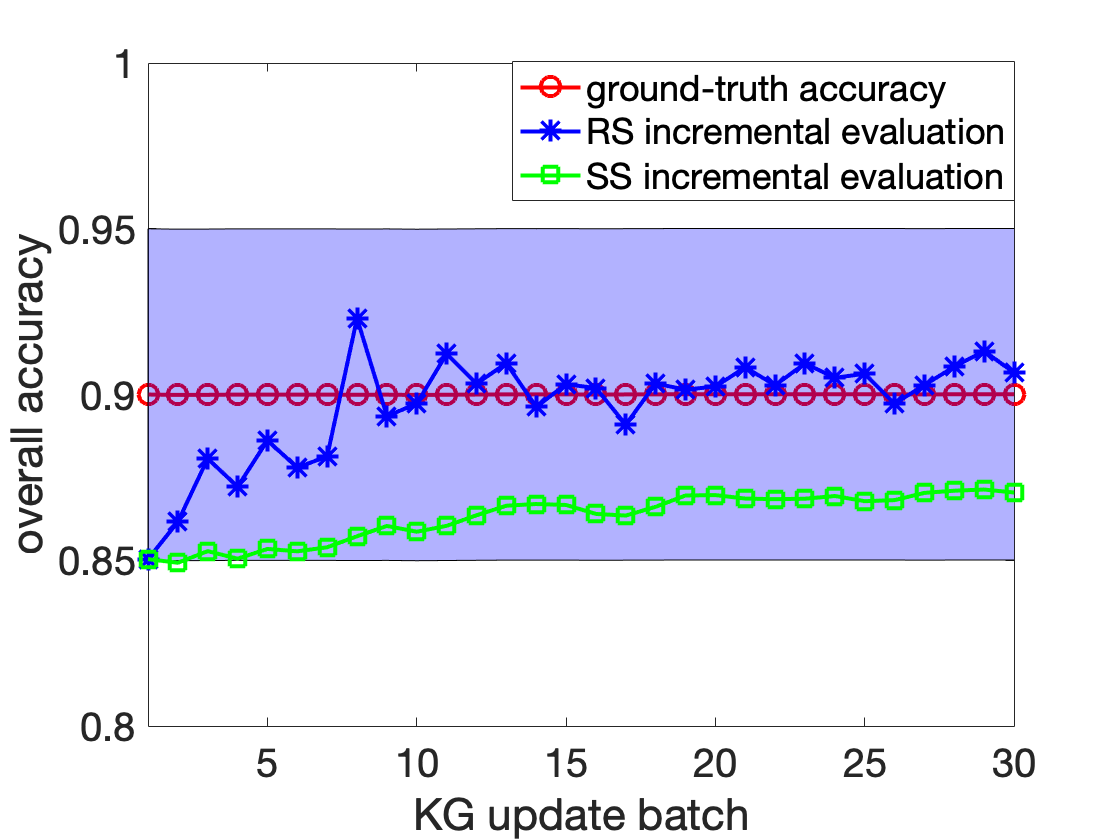}}
    \caption{\ansb{Comparing evaluation quality of incremental evaluation solutions on evolving KG with a sequence of updates: unbiasedness vs. fault tolerance. Blue ribbon in plots represents the 5\% MoE range of ground truth accuracy. }}
    \label{fig:sequence-updates}
    \vspace{-1.5em}
\end{figure}

\subsubsection{Sequence of Updates}\label{sec:expr:sequence-update}
\ansb{
In practice, we are more likely to continuously monitor the KG accuracy as it evolves. In this section, we consider the scenario of applying a sequence of updates to the base KG and compare the performance of RS and SS.
Suppose 30 update batches with similar sizes (about 10\% of the base KG) and 90\% accuracy are sequentially applied to the base KG, and an evaluation is required after each batch.
Figure~\ref{fig:sequence-updates}-1 demonstrates that both RS and SS provide unbiased estimation at every state of evolving KG.

However, if by chance the initial accuracy estimation on the base KG is significantly off, RS corrects the estimation faster than SS in a sequence of updates,  because SS reuses all samples collected in the base KG, while RS stochastically refreshes the pool with samples from the updates. This is demonstrated in Figure~\ref{fig:sequence-updates}-2 and Figure~\ref{fig:sequence-updates}-3, which show two specific runs of evolving KG evaluation starting with an initial over-estimation and under-estimation respectively. 
It is clear that SS hardly recovers from the bad estimation at the beginning, while RS is more fault-tolerant, quickly jumping away from the bad estimation and converging to the ground-truth after 5 to 10 batches of updates.

Based on the experimental results, we recommend applying RS when the quality of KG is fairly high, update is frequent and stable over time, and the update size is small comparing with the base; this is because RS cost is comparable to SS in such cases, but RS avoids the complexity of recording update history and is more robust. In other cases, we recommend SS as it may significantly reduce evaluation cost.

}

%% file: 8-Related-work.tex
\vspace{-0.5em}
\section{Related Work}\label{sec:related-work}
As discussed earlier, SRS is a simple but prevalent method for KG accuracy evaluation.
Beyond SRS, Ojha et al. \cite{ojha2017kgeval} were the first to systematically tackle the problem of efficient accuracy evaluation of large-scale KGs. 
One key observation is that, by exploring dependencies (i.e., type consistency, Horn-clause coupling constraints~\cite{bragg2013crowdsourcing,mitchell2018never,lao2011random} and positive/negative rules~\cite{ortona2018robust}) among triples in KG, one can propagate the correctness of evaluated triples to other non-evaluated triples.
The main idea of their solution is to select a set of triples such that knowing the correctness of these triples could infer correctness for the largest part of KG.
Then, KG accuracy is estimated using all labelled triples.
Their inference mechanism based on Probabilistic Soft Logic~\cite{brocheler2012probabilistic} could significantly save manual efforts on evaluating triples' correctness.
However, there are some issues in applying their approach to our setting.
First, the inference process is probabilistic and might lead to erroneous propagations of true/false labels.
Therefore, it is difficult to assess the bias introduced by this process into the accuracy estimation.
Second,  KGEval relies on expensive\footnote{According to \cite{ojha2017kgeval}, it takes more than 5 minutes to find the next triple to be manually evaluated, even on the tiny KGs with less than 2,000 triples.} (machine time) inference mechanism, which does not scale well on large-scale KGs.
Finally, they do not address accuracy evaluation for evolving KGs.
We summarize the comparison between these existing approaches in Table~\ref{tab:summary}.

Accuracy evaluation on KGs is also closely related to error detection and fact validation on KGs or ``Linked Data''~\cite{heath2011linked}. 
Related work includes numerical error detection~\cite{li2015probabilistic}, error detection through crowdsourcing~\cite{acosta2016detecting},
matchings among multiple KGs~\cite{liu2017measuring}, fact validation through web-search~\cite{gerber2015defacto}, etc.
However, previous work mentioned above all have their own limitations, and have so far not been exploited for efficient KG accuracy evaluation.

Another line of related work lies in data cleaning~\cite{chu2016data}, where sampling-based methods with groundings in statistical theory are used to improve efficiency.
In~\cite{marchant2017search}, the authors designed a novel sequential sampler and a corresponding estimator to provide efficient evaluations (F-measure, precision and recall) on the task of entity resolution~\cite{christen2007quality}.
The sampling framework sequentially draws samples (and asks for labelling) from a biased instrumental distribution and updates the distribution on-the-fly as more samples are collected, in order to quickly focus on unlabelled items providing more information.
Wang et al.~\cite{wang2014sample} considered combining sampling-based approximate query processing~\cite{hellerstein1997online} with data cleaning, and proposed a sample-and-clean framework to enable fast aggregate queries on dirty data.
Their solution takes the best of both worlds and provides accurate query answers with fast query time.
However, the work mentioned above did not take advantage of the properties of the annotation cost function that arise in practice in our setting---they focused on reducing the number of records to be labelled or cleaned by human workers, but ignored opportunities of using clustering to improve efficiency.
\begin{table}
    \centering
    \caption{Summary of existing work on KG accuracy evaluation.}
    \label{tab:summary}
    \begin{tabular}{c| c c |c}
    \hline
        \backslashbox{Property}{Method} & SRS & KGEval &  Ours \\\hline
        Unbiased Evaluation & \cmark & \xmark &   \cmark \\
        Efficient Evaluation &  \xmark & \cmark & \cmark \\
        \makecell{Incremental Evaluation\\ on Evolving KG} & \xmark &  \xmark&  \cmark\\\hline
    \end{tabular}
\end{table}

